\documentclass[aps,notitlepage,oneside,twocolumn,superscriptaddress]{revtex4-1}
\usepackage{graphicx}
\usepackage{xcolor}
\usepackage{amssymb}
\usepackage{amsmath}

\begin{document}
\title{Silhouettes of invisible black holes} 
\author{V. I. Dokuchaev}\thanks{dokuchaev@inr.ac.ru}
\affiliation{Institute for Nuclear Research of the Russian Academy of Sciences, prospekt 60-letiya Oktyabrya 7a, Moscow 117312, Russia}
\affiliation{Moscow Institute of Physics and Technology, 9 Institutskiy per., Dolgoprudny, Moscow Region 141700, Russia}
\author{N. O. Nazarova}\thanks{nnazarov@sissa.it}
\affiliation{Scuola Internazionale Superiore di Studi Avanzati (SISSA), Via Bonomea 265, 34136 Trieste (TS) Italy}
\affiliation{International Centre for Theoretical Physics (ICTP), Strada Costiera 11, 34151 Trieste (TS) Italy}

\date{\today}

\begin{abstract}
In general relativity, isolated black holes are invisible due to an infinitely large redshift of photons propagating from the event horizon to the remote observer. However, the dark shadow (silhouette) of a black hole can be visible on the background of matter radiation lensed by the gravitational field of black holes. The black hole shadow is the celestial sphere projection of the cross section of photon capture by the black hole. If the illuminating background is far behind the black hole (at a distance much greater than the event horizon radius), a classic black hole shadow of a maximal size can also be observed. A minimal-size shadow can be observed if the same black hole is illuminated by the inner part of the accretion disk adjacent to the event horizon. In this case, the shadow of an accreting black hole is a lensed image of the northern or southern hemisphere of the event horizon, depending on the orientation of the black hole spin axis. A dark silhouette of the southern hemisphere of the event horizon is seen in the first image of the supermassive black hole M87* presented by the Event Horizon Telescope. The brightness of accretion matter is much higher than the corresponding one of the usual astrophysical stationary background in the form of numerous stars or extensive hot gas clouds. For this reason, it is improbable that a black hole shadow can be observed in the presence of very luminous accretion matter.
\end{abstract}
\maketitle \tableofcontents

\section{Introduction}
\label{intro}

The black holes predicted by general relativity (Einstein's theory of gravity) are known to be invisible objects by their nature. Strictly speaking, the event horizon of any black hole represents a surface formed by geodesic photon trajectories that do not end at spatial infinity  \cite{hawkingellis}. In simple terms, the photons forming (generating) the event horizon cannot reach the remote observer. Nevertheless, in real astrophysical conditions, astrophysical black holes can be seen as dark shadows (dark silhouettes) on a radiation background. In cosmic situations, it can be either a bright background of luminous stars and extended gas clouds located far behind the black hole or a hot matter falling onto the black hole and radiating near its event horizon.

In recent years, studies of black hole images widely use the terminology taken from photography -- photographed objects (black holes), object image, its shadow, silhouette, silhouette contour. In usual photography, the photographed object can be visible or invisible. However, in the case of black holes, we are dealing with an object that is invisible by itself if described by general relativity (GR). In the image of an astrophysical black hole, one can see only its dark shadow (dark silhouette) if an external radiation source illuminates the black hole. The title of this paper emphasizes precisely this feature of black hole images. In particular, inner parts of silhouettes of white holes and wormholes can be visible, unlike totally dark black hole silhouettes.

Astrophysical observations of the most likely candidates, such as supermassive objects SgrA* in the center of our Galaxy and M87* in the center of galaxy M87, can directly probe the existence of black holes.

Direct proof of the existence of black holes is the detection of an object with the total mass contained inside the event horizon. Such direct evidence was first presented in April 2019 when the international EHT (Event Horizon Telescope) collaboration published an image of the supermassive black hole in galaxy M87 with a record high angular resolution of the order of several microarcseconds \cite{EHT1,EHT2,EHT3,EHT4,EHT5,EHT6}. For the mass $6 \times  10^9M_\odot$, the event horizon of this black hole has exactly this angular size. The obtained unique image qualitatively consists of a bright asymmetric ring, which is interpreted as the glow of an accretion disk, and a dark central part, which is interpreted as an observed black hole silhouette.

The M87 black hole image presented by the EHT collaboration is consistent with the black hole silhouette shape predicted by GR. Hence, this is the first direct evidence of the existence of black holes in the universe. Strictly speaking, all presently known astrophysical observations of black hole candidates, even including the successful detection of several gravitational wave events that can be explained only by stellar-mass binary black hole coalescences, provide circumstantial evidence of the existence of black holes.

The most crucial goal of the EHT observations also includes the imaging of the nearby supermassive black hole SgrA* in the galactic center \cite{Fish16,Lacroix13,Kamruddin,Johannsen16,Johannsen16b,Broderick16,Chael16,Kim16,Roelofs17,Doeleman17}. The supermassive black hole in the center of the Milky Way with the mass $M=(4.3 \pm 0.3)\times 10^6M_\odot$ \cite{Gheetal08,Gillessen09,Giletal09-2,Meyer12,Johannsen12} has been the focus of the most thorough research \cite{Baade46,Becklin68,Eckart96,Dokuchaev77,Dokuchaev89,Allen90,Dokuchaev91,Dokuchaev91b,Manko92,Lo93,Backer93,Haller96,Ghez98,Backer99,Reid99,Baganoff99,Falcke00b,NovikovFrolov01,Baganoff01,Hornstein02,Genzel03,Aschenbach04,YusefZadeh06,Marrone08,Ghez08,Doeleman08,Doeleman09,DoddsEden09,Broderick09,Sabha10,Dexter10,Paolis11,Broderick11,Neilsen13,Zakharov13,Fish14,Gwinn14,Johnson14,Dokuch14,Moscibrodzka14,Bower15,Johnson15,Chatzopoulos15,FizLab,Rauch16,Zakharov16,Becerril16,Giddings16,Johannsen16c,OrtizLeon16,Parsa17,Capellupo17,Shiokawa17,Johnson17,Eckart17,Abdujabbarov17,Ponti17,Zajacek18,Abuter18,Zakharov18a,Zakharov18b,Zakharov18c,Zhu19,Izmailov19,Zakharov19,TuanDo19,Do19,Giddings19,Dai19,Moriyama19}. First of all, this supermassive black hole is the closest `napping' or `sleeping' quasar with deficient activity, which is hopefully transparent for observations. Second, the technological level of the EHT project and related BlackHoleCam (Black Hole Camera) \cite{Goddi17} and GRAVITY \cite{GRAVITY18,GRAVITY19} projects enables taking black hole images with an angular resolution corresponding to their event horizons \cite{Mielnik62,Synge63,Bardeen73,Young76,Chandra,Falcke00,Takahashi04,Kardashev07,Falcke13,Li14,Inoue14,Cunha15,Abdujabbarov15,Younsi16}.

The observed shapes of black hole images depend on the matter distribution around the black hole and near its projection on the sky. The maximum black hole shadow size is observed in the case of a remote luminous background near the black hole sky projection. The dark shadow image is a sky projection of the photon capture (absorption) cross section in the black hole gravitational field. A bright background can be created by extended hot gas clouds and bright stars orbiting around the black hole, as well as by steady parts of an accretion disk outside the light spheres (the spherical photon orbits).

In addition to the remote bright background, the gas accreting onto a black hole and radiating at the black hole horizon can be observed. In Section~\ref{infall}, we show that the shadow of a black hole illuminated by an accretion disk has a smaller size than in the case of background illumination.

The shadow of an accreting black hole is shaped by high-redshifted photon trajectories near the event horizon. A detailed image of the shadow from an accreting black hole is sensitive to the angular resolution of the telescope. The shadow of an accreting black hole represents a lensed image of its event horizon, and it is therefore natural to call this shadow the silhouette of the black hole event horizon. For a thin accretion disk, the black hole event horizon silhouette turns out to be a lensed image of the northern or southern hemisphere of the event horizon globe, depending on the orientation of the black hole spin. The contour of this silhouette is a lensed image of the equator on the event horizon globe. In the image of the supermassive black hole M87* obtained by the EHT, the dark silhouette of the event horizon of this black hole represents a lensed image of the southern hemisphere of the event horizon.

In this paper, the black hole shadow with maximal size formed on a remote background is referred to as the classical black hole shadow. This type of black hole shadow is the most actively studied in recent publications. We demonstrate that for an accreting black hole, a shadow with minimal diameter is observed. In this case, the shadow is the silhouette of the event horizon itself.

The enormous energy released from an accreting black hole provides a high luminosity of the accretion disk compared to the background radiation from gas and stars surrounding the black hole. In other words, in real astrophysical conditions, the accretion disk producing the event horizon image of a black hole is much brighter than the remote background creating the classical black hole shadow. For this reason, using even very high angular resolution observations, it is much more likely to observe the silhouette of the event horizon of an accreting black hole than the classical black hole shadow. As a result, the conventional black hole shadow is difficult to detect either because of the low accretion activity of the black hole or due to the shadow blurring by powerful accretion disk emission.

The forthcoming EHT upgrade will enable experimental studies of black holes and shapes of their images to test GR and its modification in the strong-field limit \cite{deVries00,Schnittman06,Shatskiy08,Bambi09,Frolov09,Tamburini11,Vincent11,Amarilla12,Johannsen13,Babichev13,Amarilla13,Zakharov14,Wei15,Abd15,Nucamendi15,Nucamendi16,Cunha16,Abdujabbarov16,CliffWill17a,Cunha17,CliffWill17b,Amarilla17,Mureika17,DokEr15,Amir18,Lan18,Wang18,Lamy18,Mizuno18,Repin18,Wei19,Blackburn19,Meierovich19,Abdikamalov19,Zhu19b,Tian19,Davoudiasl19,Konoplya19b,Hess19,Rummel20,Alexeyev20}. Prospective instruments for future studies are space interferometers with nanoarcsecond angular resolution \cite{Kardashev14,Roelofs19,Palumbo19}.

We note that in real cosmic conditions, black hole masses increase by accretion of the surrounding matter or due to coalescence with other black holes. The outer capture boundary of a black hole with changing mass is called the apparent local horizon. In contrast, the global event horizon is determined by the full history of the black hole in the universe and is determined by its ultimate mass (see, e.\,g., \cite{Wald} for more details). Below, we do not distinguish between the apparent horizon and the event horizon, ignoring the black hole mass change.

	Traditionally it is thought that the event horizon of a black hole is invisible and it is impossible to reconstruct its image. In Section 6.3, we show that the gravitational lensing of the emitting matter falling into a black hole outside its equatorial plane offers a fundamental possibility of constructing the lensed image of the entire event horizon of the black hole. This image is projected onto the sky inside the black hole shadow. Here, the image of the event horizon is a lensed sky of the total surface of the event horizon and not of its frontal size only. Black holes turn out to be unique objects in the universe that can be seen simultaneously from all sides.

In the future, the black hole shadow and image of its event horizon can be detected simultaneously in observations with different telescopes with required angular resolution. Namely, the shadow of a black hole on a distant background can be measured, e.\,g., in the X-ray or near-infrared (IR) band in observations of extended background sources and lensed images of ordinary stars and neutron stars behind the black hole. At the same time, a remote observer can take a picture of the event horizon by detecting high-redshifted photons produced near the event horizon by hot matter accreting onto the black hole.

\section{Rotating Kerr black hole}
\label{Kerr0}

It is convenient to present the famous Kerr metric  \cite{Kerr}, which is the exact solution of Einstein's equations for a rotating black hole, in the Boyer--Lindquist coordinates $(t,r,\theta,\phi)$  \cite{BoyerLindquist} in the standard form valid for any stationary axially symmetric asymptotically flat space--time \cite{Bardeen70,Bardeen70b,BPT,mtw,Galtsov}:  
\begin{equation}
ds^2=-e^{2\nu}dt^2+e^{2\psi}(d\phi-\omega dt)^2+e^{2\mu_1}dr^2+e^{2\mu_2}d\theta^2.
\label{metric}
\end{equation}
This standard metric represents the Kerr solution with 
\begin{eqnarray}
e^{2\nu}&=&\frac{\Sigma\Delta}{A}, \quad e^{2\psi}=\frac{A\sin^2\theta}{\Sigma}, 
\nonumber \\
e^{2\mu_1}&=&\frac{\Sigma}{\Delta}, \quad e^{2\mu_2}=\Sigma, \quad
\omega=\frac{2Mra}{A},
\label{omega}
\end{eqnarray}
where
\begin{eqnarray}
\Delta &= & r^2-2Mr+a^2, \label{Delta} \\
\Sigma &=& r^2+a^2\cos^2\theta,  \label{Sigma} \\
A&=& (r^2+a^2)^2-a^2\Delta\sin^2\theta. 
\label{A}
\end{eqnarray}
$M$ is the black hole mass, $a = J/M$ is the specific angular momentum (spin) of the black hole, and $\omega$ is the so-called frame-dragging angular velocity. Geometrical units $G = 1$ and $c = 1$ are used. To simplify the formulas below, we frequently use the dimensionless length $r\Rightarrow r/M$ and time $t\Rightarrow t/M$, which is equivalent to the condition $M = 1$. In other words, we express dimensionless radial distances in units of $GM/c^2$ and the corresponding time intervals in units of $GM/c^3$. Accordingly, we use the dimensionless black hole spin $a=J/M^2\leq1$ and assume that $a \leq1$ without loss of generality.

The most general laws of dynamics (or thermodynamics) of stationary axially symmetric black holes were formulated in the seminal paper \cite{bch}.

The event horizon of a Kerr black hole in the Boyer--Lindquist coordinates is a sphere with the radius
\begin{equation}
r_{\rm h}=M+\sqrt{M^2-a^2},
\label{rh}
\end{equation}
It is the larger root of the quadratic equation  $\Delta=0$. The event horizon exists only for $a\leq M$. For $a>M$, there is no event horizon, and Kerr metric (\ref{metric})  describes a naked singularity. For $a=0$, the Kerr metric coincides with the Schwarzschild metric for a static spherically symmetric black hole. We emphasize that, strictly speaking, the event horizon of a Kerr black hole has a spherical shape only in the topological sense, because the Gaussian curvature of the event horizon surface is not constant and depends on the polar angle $\theta$ \cite{Smarr73,Sharp81}. The purely spherical form of the event horizon of a Kerr black hole in the Boyer--Lindquist coordinates is the salient feature of this unique reference frame.

A remarkable property of the frame-dragging angular velocity $\omega$ in Eqn (\ref{omega}) is its independence on the angle $\theta$  at the black hole event horizon: 
\begin{equation}
\omega(r_{\rm h})=\Omega_{\rm h}=\frac{a}{2M(M+\sqrt{M^2-a^2})}.
\label{Omegah}
\end{equation}

The angular velocity $\Omega_{\rm h}$  is called the angular velocity of the black hole horizon. Thus, the event horizon in the Boyer--Lindquist coordinates rotates as a rigid body!

\section{Locally nonrotating frames}
\label{LNRF}

In any stationary axially symmetric asymptotically flat space--time, it is convenient to use orthonormal locally nonrotating reference frames (LNRFs) \cite{BPT,Bardeen70,Bardeen70b} in which observers are moving along the world lines $r=const$, $\theta=const$, $\varphi=\omega t+const$. In the Kerr metric, the frame-dragging angular velocity $\omega$ is defined by Eqn~(\ref{omega}). Physical observers in the LNRF `rotate' together with the black hole, and, for them, all physical processes in the Kerr metric appear in the most natural way, as opposed to how they look in other frames \cite{BPT}.

The basis vectors of an orthonormal tetrade connected to the physical observers in the LNRF have the form \cite{BPT}: 
\begin{eqnarray}
{\mathbf e}_{(t)}&\!=\!&e^{-\nu}\left(\frac{\partial}{\partial t}
\!+\!\omega\frac{\partial}{\partial\varphi}\right)\!=\!
\left(\frac{A}{\Sigma\Delta}\right)^{1/2}\!\frac{\partial}{\partial t}\!+\!
\frac{2Mar}{(A\Sigma\Delta)^{1/2}}\frac{\partial}{\partial\varphi}, \label{et} 
\nonumber \\
{\mathbf e}_{(r)}&=&e^{-\mu_1}\frac{\partial}{\partial r} =
\left(\frac{\Delta}{\Sigma}\right)^{1/2}\frac{\partial}{\partial r}, \label{er} 
\nonumber \\
{\mathbf e}_{(\theta)}&=&e^{-\mu_2}\frac{\partial}{\partial\theta} =
\frac{1}{\Sigma^{1/2}}\frac{\partial}{\partial\theta}, \label{etheta} 
\nonumber \\
{\mathbf e}_{(\varphi)}&=&e^{-\psi}\frac{\partial}{\partial\varphi} =
\left(\frac{\Sigma}{A}\right)^{1/2}\!\frac{1}{\sin\theta}
\frac{\partial}{\partial\varphi}.\label{evarphi} 
\end{eqnarray}

The corresponding basis differential one-forms (or covariant basis vectors) are expressed as 
\begin{eqnarray}
{\mathbf e}^{(t)}&=&e^{\nu}{\mathbf d}t=
\left(\frac{\Sigma\Delta}{A}\right)^{1/2}\!{\mathbf d}t, \label{et2} 
\nonumber \\
{\mathbf e}^{(r)}&=&e^{\mu_1}{\mathbf d}t=
\left(\frac{\Sigma}{\Delta}\right)^{1/2}\!{\mathbf d}r, \label{er2}  
\nonumber \\
{\mathbf e}^{(\theta)}&=&e^{\mu_2}{\mathbf d}t=
\Sigma^{1/2}\!{\mathbf d}\theta, \label{etheta2}  
\nonumber \\
{\mathbf e}^{(\varphi)}&=&-\omega e^{\psi}{\mathbf d}t+e^{\psi}{\mathbf d}\varphi
\nonumber \\
&=&-\frac{2Mar\sin\theta}{(\Sigma A)^{1/2}}{\mathbf d}t+
\left(\frac{A}{\Sigma}\right)^{1/2}\!\!\sin\theta{\mathbf d}\varphi. \label{evarphi2} 
\end{eqnarray}
Equations~(\ref{evarphi})   and (\ref{evarphi2}) define the components  $e^\mu_{\phantom{1}(i)}$ and $e_\mu^{\phantom{1}(i)}$ of basis vectors in the LNRF: 
\begin{equation}
{\mathbf e}_{(i)}=e^\mu_{\phantom{1}(i)}\frac{\partial}{\partial x^\mu} \quad
\mbox{and} \quad 
{\mathbf e}^{(i)}=e_\mu^{\phantom{1}(i)}{\mathbf d}x^\mu.
\label{basis}
\end{equation}
In particular, the standard transformations of the second rank tensors with these components have the form 
\begin{equation}
J_{(a)(b)}=e^\mu_{\phantom{1}(a)}e^\nu_{\phantom{1}(b)}J_{\mu\nu}, \quad
J_{\mu\nu}=e_\mu^{\phantom{1}(a)}e_\nu^{\phantom{1}(b)}J_{(a)(b)}.
\label{basis}
\end{equation}
In Section~\ref{infall}, we use the LNRF to calculate the energy shift of a photon due to gravitational redshift and the Doppler effect associated with lensing in the black hole gravitational field.

\section{Equations of motion for test particles}

Integrals of motion fully determine the trajectory of a massive test particle in the Kerr metric (disregarding gravitational radiation). These are the test particle mass $\mu$, the total energy $E$, the azimuthal angular momentum $L$, and the Carter constant $Q$ related to the nonazimuthal part of the particle angular momentum \cite{Carter68,Chandra}. In particular, for $Q = 0$, the motion of the particle is bounded to an orbital plane. Using these integrals of motion, Carter \cite{Carter68} obtained first-order differential equations for test particles. Carter's remarkable equations have the form \cite{Carter68,deFelice,Chandra,BPT,mtw,Galtsov} 
\begin{eqnarray}
\Sigma\frac{dr}{d\tau} &=& \pm \sqrt{R(r)}, \label{rmot} \\
\Sigma\frac{d\theta}{d\tau} &=& \pm\sqrt{\Theta(\theta)}, \label{thetamot} \\
\Sigma\frac{d\varphi}{d\tau} &=& L\sin^{-2}\theta+a(\Delta^{-1}P-E),
\label{phimot} \\
\Sigma\frac{dt}{d\tau} &=& a(L-aE\sin^{2}\theta)+(r^2+a^2)\Delta^{-1}P,
\label{tmot}
\end{eqnarray}
where $\tau$ is the proper time of a massive particle or an affine parameter along the trajectory of a massless particle ($\mu=0$). In these equations, the effective radial potential determines the radial and polar motion,
\begin{equation}
R(r) = P^2-\Delta[\mu^2r^2+(L-aE)^2+Q],
\label{Rr} 
\end{equation}
where $P=E(r^2+a^2)-a L$ and the effective polar potential is 
\begin{equation}
\Theta(\theta) = Q-\cos^2\theta[a^2(\mu^2-E^2)+L^2\sin^{-2}\theta].
\label{Vtheta} 
\end{equation}
Notably, zeros of these effective potentials determine the turning points $dR/d\tau=0$ and $d\Theta/d\tau=0$  in the radial and polar directions.

Trajectories of massive particles ($\mu\neq0$) in the Kerr metric are  determined by three  parameters (constants of motion): $\gamma=E/\mu$, $\lambda=L/E$ and $q=\sqrt{Q}/E$. The corresponding photon trajectories (light or null geodesics) are determined by two constant of motion:   $\lambda=L/E$ and $q=\sqrt{Q}/E$ (we ignore possible photon trajectories with $Q<0$ in the Kerr metric, because they do not reach the remote observer).

Differential equations (\ref{rmot})--(\ref{tmot}) can be represented in the integral form  \cite{Carter68,BPT,Chandra,mtw}, convenient for numerical integration: 
\begin{equation}\label{eq2425}
\int\limits_{\cal C}\frac{dr}{\sqrt{R(r)}}
=\int\limits_{\cal C}\frac{d\theta}{\sqrt{\Theta(\theta)}},
\end{equation}
\begin{equation}\label{eq2425b}
\tau=\int\limits_{\cal C}\frac{r^2}{\sqrt{R(r)}}\,dr
+\int\limits_{\cal C}\frac{a^2\cos^2\theta}{\sqrt{\Theta(\theta)}}\,d\theta,
\end{equation}
\begin{equation}
\phi=\int\limits_{\cal C}\frac{aP}{\Delta\sqrt{R(r)}}\,dr
+\int\limits_{\cal C}\frac{L-aE\sin^2\theta}{\sin^2\theta\sqrt{\Theta(\theta)}}\,d\theta, \label{eq25bbb} 
\end{equation}
\begin{equation}
t=\int\limits_{\cal C}\frac{(r^2+a^2)P}{\Delta\sqrt{R(r)}}\,dr
+\int\limits_{\cal C}a\frac{L-aE\sin^2\theta}{\sqrt{\Theta(\theta)}}\,d\theta, \label{eq25tt} 
\end{equation}
where the effective potentials $R(r)$ and $\Theta(\theta) $ are defined by formulas (\ref{Rr}) and (\ref{Vtheta}). Integrals in (\ref{eq2425})--(\ref{eq25tt}) are contour integrals of the first kind along the particle trajectory $C$. The most important feature of these integrals is their monotonic increase along the particle path. 
Formally, this means that the integrands do not change sign when passing through the radial and angular turning points. In particular, the contour integrals in (\ref{eq2425}) reduce to ordinary integrals in the absence of the radial $r$ and angular $\theta$ turning points along the trajectory: 
\begin{equation}\label{eq24a}
\int^{r_s}_{r_0}\frac{dr}{\sqrt{R(r)}}
=\int_{\theta_0}^{\theta_s}\frac{d\theta}{\sqrt{\Theta(\theta)}},
\end{equation}
Here,  $r_s$ and $\theta_s$ the radius and latitude of the initial point of the trajectory (for example, the point of photon emission); $r_0\gg r_{\rm h}$ and $\theta_0$  are the coordinates of the endpoint of the trajectory (for example, the detection point by a remote observer). If the trajectory has one turning point $\theta_{\rm min}(\lambda,q)$ (the extremum of the polar potential $\Theta(\theta)$), the contour integrals in (\ref{eq2425}) are expressed through ordinary integrals: 
\begin{equation}\label{eq24b}
\int_{r_s}^{r_0}\frac{dr}{\sqrt{R(r)}}
=\int_{\theta_{\rm min}}^{\theta_s}\frac{d\theta}{\sqrt{\Theta(\theta)}}
+\int_{\theta_{\rm min}}^{\theta_0}\frac{d\theta}{\sqrt{\Theta(\theta)}}.
\end{equation}
Correspondingly, the contour integrals in (\ref{eq2425})  for trajectories with two turning points, $\theta_{\rm min}(\lambda,q)$  and $r_{\rm min}(\lambda,q)$ (the extremum of the radial potential $R(r)$), are also expressed through ordinary integrals:
\begin{equation}\label{eq24c}
\int_{r_{\rm min}}^{r_s}\!\!\frac{dr}{\sqrt{R(r)}}
+\int_{r_{\rm min}}^{r_0}\!\!\frac{dr}{\sqrt{R(r)}}
=\!\!\int_{\theta_{\rm min}}^{\theta_s}\!\!\frac{d\theta}{\sqrt{\Theta(\theta)}}
+\!\!\int_{\theta_{\rm min}}^{\theta_0}\!\!\frac{d\theta}{\sqrt{\Theta(\theta)}}.
\end{equation}
Integral equations (\ref{eq2425})--(\ref{eq25tt}) for trajectories with additional turning points can be similarly expressed through ordinary integrals.

In the Kerr metric, shapes of trajectories remain the same during the simultaneous change of the sign of the time $t$ and the angular velocity of metric rotation $\omega$ (i.\,e., after the flip of the black hole spin). For example, the trajectory of a particle falling into a black hole coincides with the trajectory of the same particle (i.\,e., of the particle with the same integrals of motion) ejected from a white hole. We also note that in the Kerr metric, test particle trajectories around a black hole are the same for a white hole if we change the sign of time $t$.

\section{Features of particle trajectories }

\subsection{Inevitable rotation in the ergosphere}

 The most striking feature of the gravitational field of a rotating Kerr black hole outside the horizon is the existence of the `ergosphere' -- a region in which physical observers cannot be `at rest' (with $r,\theta,\varphi=const$) relative to a remote observer. The inner boundary of the ergosphere coincides with the black hole horizon $r_{\rm h}=1+\sqrt{1-a^2}$. The outer boundary of the ergosphere (the static limit) $r_{\rm ES}(\theta)$ is 
\begin{equation}\label{ES}
r_{\rm ES}(\theta) =1+\sqrt{1-a^2\cos^2\theta}.
\end{equation}
Inside the ergosphere, all material objects, including photons, are dragged into rotation around the black hole with the angular velocity $\dot\varphi=d\phi/dt\propto a$, which tends to the horizon angular velocity $\Omega_{\rm h}$ from (\ref{Omegah}) in approaching the horizon. Inside the ergosphere, there are geodesics with negative total energy, $E<0$, falling into the black hole. Such geodesics are responsible for the practicability of the `Penrose mechanism' of extraction of rotational energy from the black hole \cite{mtw,Penrose}.

\subsection{Winding on the event horizon}
\begin{figure}
	\centering 
	\includegraphics[angle=0,width=0.45\textwidth]{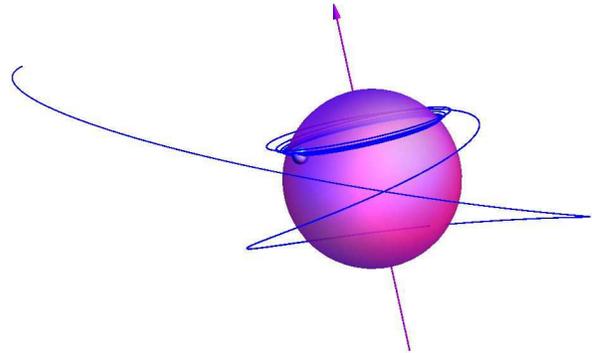}
	\caption{(Color online.) $3D$ trajectory of a massive test particle falling into a black hole with spin $a=0.998$. The particle trajectory starts at the radius $r(0)=4.4$. In approaching the black hole, the particle (the small blue ball) starts winding up on the event horizon with a constant angular velocity $\Omega_{\rm h}$. The purple sphere is the black hole horizon. The purple axis marks the black hole rotational axis.}
	\label{fig1}
\end{figure}

\begin{figure}
	\centering 
	\includegraphics[angle=0,width=0.45\textwidth]{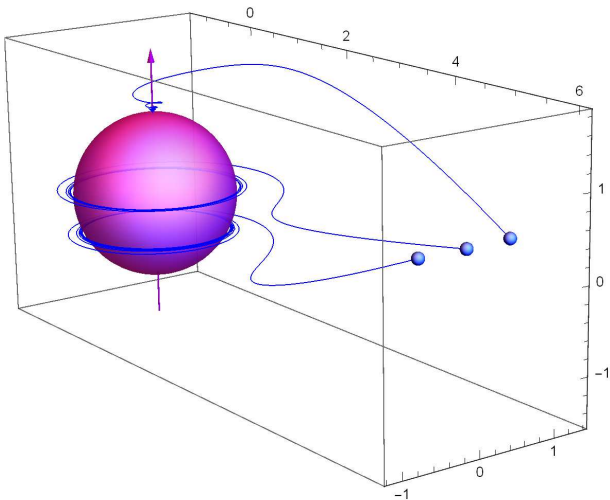}
	\caption{(Color online.) $3D$ trajectories of massive test particles falling into a black hole with spin $a=0.998$ near the north pole of the event horizon, near its equator, and in the southern hemisphere. In approaching the black hole, all particles start winding up on the event horizon with a constant angular velocity $\Omega_{\rm h}$.}
	\label{fig2}
\end{figure}

\begin{figure}
	\centering 
	\includegraphics[angle=-90,width=0.45\textwidth]{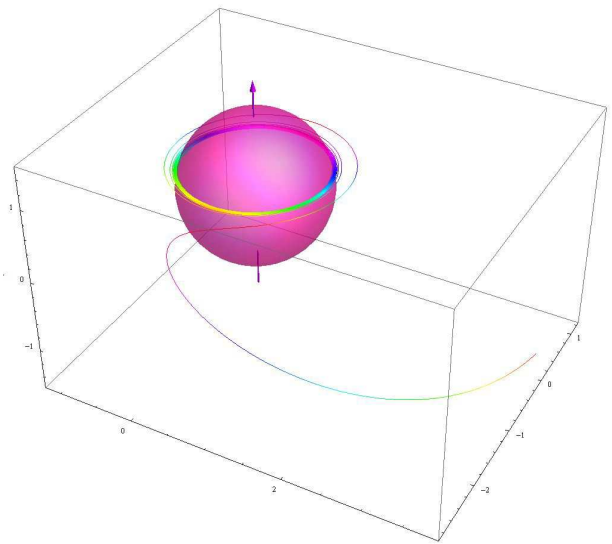}
	\caption{(Color online.) $3D$ trajectory of a photon falling with a negative azimuthal angular momentum. Outside the ergosphere, the photon moves with a negative angular velocity relative to the remote observer, $d\varphi/dt<0$. After entering the black hole ergosphere, the photon angular velocity changes sign and becomes positive, $d\varphi/dt>0$. Moreover, in approaching the black hole event horizon, the photon is dragged into its rigid-body rotation, $d\varphi/dt\to\Omega_{\rm h}=const$.}
	\label{fig3}
\end{figure}

The second remarkable feature of the gravitational field of a Kerr black hole is the azimuthal winding of particle trajectories on the event horizon when falling into the black hole with $a\neq0$. As $r\to r_{\rm h}$, the angular velocity of the trajectory winding of all particles falling into the black hole (including photons) in the Boyer--Lindquist coordinates tends to the constant angular velocity of the event horizon $\Omega_{\rm h}$ in (\ref{Omegah}), which is independent of the polar angle $\theta$. 
In other words, all particles falling into the black hole by approaching the horizon are dragged into its solid-body-like rotation, $d\varphi/dt\to\Omega_{\rm h}=const$ (see the illustration of this effect in Figs ~\ref{fig1}--\ref{fig3}). The orbital parameters $\lambda$ and $q$ for these $3D$ trajectories are found by numerically solving the integral equations of motion (\ref{eq2425}). The $3D$ trajectories of massive particles  ($\mu\neq0$) and photons ($\mu=0$) are computed by numerical integration of the differential equations of motion (\ref{rmot})--(\ref{tmot}).


Figure~\ref{fig1} shows the $3D$ trajectory of a massive test particle ($\mu\neq0$) with the orbital parameters $Q=0.3\,M^2\mu^2$, $E/M=0.85$, and $L\Rightarrow L/M=1.7$ falling into a black hole with the spin $a=0.998$. The trajectory starts at the radius $r(0)=4.4$. In all $3D$ pictures, we use the Boyer--Lindquist coordinates, because both the black hole horizon and trajectories of all particles can be presented most simply and visually in these coordinates. 

Figure~\ref{fig2} shows similar examples of $3D$ trajectories of massive test particles (small blue balls) falling into a black hole with the spin $a=0.998$ near the north pole of the event horizon ($\gamma=1$, $\lambda=0$, $q=1.85$), near its equator ($\gamma=1$, $\lambda=-1.31$, $q=0.13$), and in the southern horizon hemisphere ($\gamma=1$, $\lambda=-1.31$, $q=0.97$). In approaching the black hole, all particles start winding up on the event horizon with the constant angular velocity $\Omega_{\rm h}$ in (\ref{Omegah}).

Figure~\ref{fig3} shows the trajectory of a massless particle (a photon) falling into a black hole with the impact parameter $b=L/E=-6.5$ (i.\,e., with the azimuthal angular momentum directed against the black hole spin) and the Carter parameter $Q=4$. The trajectory of this photon illustrates both striking features of the gravitational field of a rotating black hole: the irresistible rotation around the black hole inside the ergosphere and the winding of the particle trajectory on the event horizon. Outside the ergosphere, this photon moves with a negative angular velocity relative to a remote observer, $d\varphi/dt<0$. When entering the ergosphere, the angular velocity of the photon motion changes sign and becomes positive, $d\varphi/dt>0$. In approaching the black hole, the photon, like massive particles, starts winding up on its event horizon with the constant angular velocity $\Omega_{\rm h}$.


\subsection{Spherical orbits}
\label{spherical}

The third feature of the gravitational field of a rotating black hole is the existence of relativistic spherical orbits of particles along which they move on a sphere with a radius $r=const$ by oscillating in the polar direction between the turning points $\theta_{\rm min}$ and $\theta_{\rm max}=\pi-\theta_{\rm min}$. Spherical orbits in the Kerr metric were studied in detail by Wilkins \cite{Wilkins} (see also \cite{Grossman12,Hod13,Liu19,Glampedakis19}). In the particular case of equatorial orbits (for $Q=0$), the spherical orbits reduce to circular orbits or rings \cite{BPT}.

\begin{figure}[h]
	\includegraphics[width=0.48\textwidth]{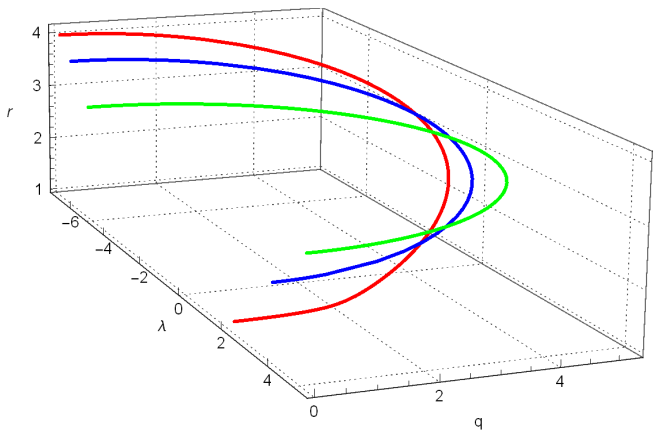}
	\caption{(Color online.) For $0<a\leq1$, radii of photon spheres depend on one of the  photon  parameters, $\lambda$ or $q$. For $a=1$  ({\color{red} red} curve), $r_{\rm ph}=1+\sqrt{2-\lambda}$, $q^2= (1+\sqrt{2-\lambda})^3(3-\sqrt{2-\lambda})$. These photon spheres exist in the radial interval $1\leq r_{\rm ph}\leq4$. For $a=0$ the photon spheres  reduce to the photon rings with the radius ({\color{green} green} semi-circle) for which $\lambda^2+q^2=27$. The {\color{blue} blue} curve corresponds to  photon spheres at $a=0.6$.}
	\label{fig4}      
\end{figure}

Spherical orbits can be found from the joint solution of the equations $R(r)=dR(r)/dr=0$, in which the effective  radial potential $R(r)$ is given by formula (\ref{Rr}), and the polar turning points are zeros of the effective polar potential $\Theta(\theta)$ from (\ref{Vtheta}). In the particular case of an extreme Kerr black hole with $a=1$, spherical orbits of massive particles (corotating with the black hole) at the radius $r=1$, which coincides with the event horizon in the Boyer--Lindquist coordinates, form a one-parameter family \cite{Wilkins}: 
\begin{equation}
\frac{2}{\sqrt{3}} \leq L\leq 2, \quad Q=\frac{3}{4}L^2-1, \quad
E=\frac{1}{2}L.
\end{equation}
As shown in Section~\ref{classic}, unstable spherical photon orbits (or photon spheres) play a crucial role in the black hole silhouette formation. In the case of a Schwarzschild black hole ($a=0$), the photon spheres reduce to photon rings with the radius $r_{\rm ph}=3$. For $a>0$, there is an infinitely large number of photon spheres with the radius depending on one of the photon orbit parameters, $\lambda$ or $q$. For $a=1$, the photon spheres are located at the radii $r_{\rm ph}=1+\sqrt{2-\lambda}$, $q^2= (1+\sqrt{2-\lambda})^3(3-\sqrt{2-\lambda})$, within the range $1\leq r_{\rm ph}\leq4$. Figure~\ref{fig4} shows the corresponding dependences of the photon sphere radii on $\lambda$ or $q$ for the black hole spins $a=0$, $a=0.6$ and $a=1$.

\subsection{Multiple images}
Lensing in the gravitational field of a black hole, generally speaking, gives rise to an infinite number of images of individual objects \cite{CunnBardeen72,CunnBardeen73,Viergutz93,RauchBlandf94,GralHolzWald19}. The Cunningham--Bardeen scheme is a convenient classification of multiple images (or light echos) \cite{CunnBardeen72,CunnBardeen73}. Each of the multiple images uniquely corresponds to the number $n$ of photon trajectory crossings of the equatorial plane of the black hole along the entire path from the source to the observer. Photons forming the direct image of the emission source (type zero or order-zero trajectories) do not cross the black hole equatorial plane along the entire path from the source to the observer. Photons of type-one trajectories (of the first light echo) intersect the black hole equatorial plane once. Correspondingly, type-$n$ light echo photons (trajectories of type $n$) cross the black hole equatorial plane $n$ times.

\begin{figure}
	\centering 
	\includegraphics[width=0.49\textwidth,origin=c,angle=0]{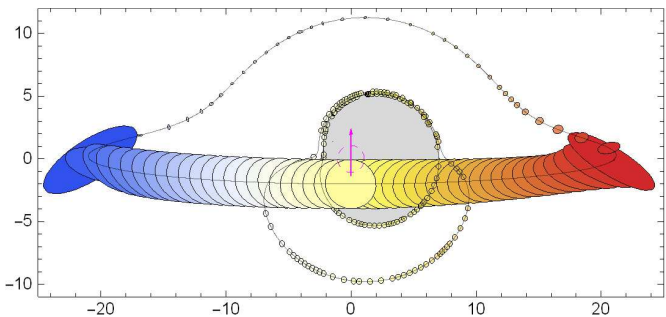}
	\caption{
	(Color online.) Direct image and the first and second light echos from a lensed compact star (radiating probe) on a circular equatorial orbit with the radius $r=20$ around the almost extreme black hole SgrA*. All multiple images of the star lie beyond the black hole shadow (grey region).}
	\label{fig5}
\end{figure}

 Figure~\ref{fig5} shows a direct image, as well as the first and second light echo, of a star rotating around an almost extreme Kerr black hole in a circular orbit of the radius $r=20$ \cite{doknaz17,doknaz18b}. Below, we model only direct images of luminous sources, because the brightness of direct images (except for the particular relative location of the source, black hole, and observer) far exceeds that of light echos \cite{CunnBardeen72,CunnBardeen73}.

\section{Shapes of black hole images } 
We consider models of lensed images of supermassive black holes SgrA* and M87* as examples. 

To numerically calculate the trajectories of massive and massless particles, we use geodesic equations in the Kerr metric (\ref{rmot})--(\ref{tmot}) and (\ref{eq2425})--(\ref{eq25tt}). To compute the brightness of lensed images of radiating matter, we use the Cunningham--Bardeen formalism \cite{CunnBardeen72,CunnBardeen73}. For the supermassive black hole SgrA* in the Milky Way center, the polar angle (latitude) of a remote observer is $\theta_0\simeq84.24^\circ\!$ , with $\cos\theta_0=0.1$ (see examples in Figs~\ref{fig4}, \ref{fig5}, \ref{fig12}--\ref{fig14}, and \ref{fig16}--\ref{fig18}).

The other supermassive black hole M87* available for EHT observations, with a mass of $(6.6 \pm 0.4)\times 10^9M_\odot$, is in the nucleus of the nearby giant elliptical galaxy M87 (NGC 4486) located in the Virgo galaxy cluster center \cite{Ho08,Gebhardt09,Gebhardt11,Walsh13}. An extended relativistic jet is observed to extend from the nucleus of this galaxy \cite{Curtis18,Rees78,Eichler83,Rees84,Begelman84,Stiavelli92,Junor95,Junor99,Matteo03,Kovalev07,Beskin10,Hada11,deGasperin12,Moscibrodzka16} at the viewing angle of $17^\circ$. Detailed long-term interferometric observations of the jet base at cm wavelengths suggest that the black hole and the surrounding accretion disk rotate clockwise \cite{Doeleman12,Broderick15,Lacroix16,Akiyama17,Walker18}, which means that the polar angle of the remote observer on Earth is $\theta_0=180^\circ-17^\circ$, which we use in our numerical modeling (see examples in Figs~\ref{fig11}, \ref{fig15}, \ref{fig19} and \ref{fig20}). 

\subsection{Classical black hole shadow on the remote background}
\label{classic}

In real astrophysical conditions, a black hole can be illuminated by hot matter, e.\,g., by extended hot gas clouds or bright stars located behind the black hole. In this case, a classical black hole shadow with the maximum size can be observed (see numerous studies \cite{Mielnik62,Synge63,Bardeen73,Young76,Chandra,Falcke00,Luminet79,Zakharov94,Beckwith05,ZakhPaoIngrNuc05,Takahashi05,Takahashi07,Bakala07,Huang07,Virbhadra09,Hioki09,Schee09,Dexter09,JohPsaltis10,Amarilla10,Nitta11,Yumoto12,Abdujabbarov13,Atamurotov13,Atamurotov13b,Wei13,Tsukamoto14,Papnoi14,Tinchev14,Kraniotis14,Ghas15,Tinchev15,Gralla15,Atamurotov15,Perlick15,Shipley16,Liu16,Yang16,Strom16,Amir16,Gralla16,Vincent16,Dastan16,Tret16,Dastan16b,Sharif16,Opatrny17,Cunha17b,Singh17,Wang17,Amir17,Strom17,Strom18,Tsupko17a,Tsupko17b,Cunha18b,BisnovatyiTsupko17,Stuchlik18,Huang18,Tsukamoto18,Bisnovatyi18,Hou18,Yan19,Gyulchev19,Kumar19,Konoplya19,Sabir19,Johnson19,Siino19,Zhang19,Shipley19,Shaikh19,Shaikh19b,Ding19,Narayan19,Goddi19,Feng19,Allahyari19,Cunha19,Konoplya19c,Jusufi19,Tsupko20,Vagnozzi20,Yu20,Li20,Chang20,Himwich20,Bakala20,Anantua20}). In the pioneering paper by Bardeen \cite{Bardeen73}, the shadow of a Kerr black hole on a remote luminous background is called the `visible boundary' of the black hole. (For more general definitions of the black hole shadow, see, e.\,g., \cite{Grenzebach14,Grenzebach15,Cunha18a}).

Photon trajectories of constant radius $r=const$ determine the observed shape of the outer boundary of the sky projection of the classical shadow of a black hole. These trajectories represent spherical orbits or photon spheres (see Section~\ref{spherical}). According to the equation of motion (\ref{rmot}), a particle remains forever in a circular orbit $r=const$ if the  orbit is the eternal turning point, i.\,e., if the condition $dr/d\tau=0$ holds not only instantaneously but at all subsequent moments of the particle's proper time $\tau$. This requirement corresponds to the conditions $R=0$ and $dR/dr=0$ for the effective potential. In the case of photons, $\tau$ is the affine parameter along the photon trajectory. Photon trajectories (orbits) of constant radius for $a\neq0$ are called spherical (or photon spheres) because, while keeping the constant radius $r=const$, the photon motion oscillates in the polar direction $\theta$. We note once again that the photon trajectories with constant radius are unstable. 

A joint solution of the equations $R(r)=0$ and $[rR(r)]'=0$ determining the shape of the black hole shadow in the parametric form $(\lambda,q)=(\lambda(r),q(r))$ (see, e.\,g., \cite{Bardeen73,Chandra}) is given by 
\begin{eqnarray}
\lambda&=&\frac{-r^3+3r^2-a^2(r+1)}{a(r-1)}, \\
q^2&=&\frac{r^3[4a^2-r(r-3)^2]}{a^2(r-1)^2}.
\label{shadow2}
\end{eqnarray}
For $a=1$, these equations are significantly simplified \cite{Bardeen73}: 
\begin{eqnarray}
\lambda&=&-r^2+2r+1, \\
q^2&=&r^3(4-r).
\label{shadow3}
\end{eqnarray}
Moreover, because of the nonuniform nature of the limit $a\to1$, the parameter $q$ at $r=1$ changes within the range $0\leq q\leq\sqrt{3}$. 

The photon trajectory parameters $\lambda=L/E$ and $q=\sqrt{Q}/E$ are related to the impact parameters $\alpha$ and $\beta$ in the sky corresponding to photons detected by a remote observer located at a given radius $r_0\gg r_{\rm h}$ (i.\,e., almost at infinity), latitude $\theta_0$, and azimuth $\varphi_0$ (see \cite{Bardeen73,CunnBardeen72,CunnBardeen73} for more details): 
\begin{equation}
\alpha =-\frac{\lambda}{\sin\theta_0}, \quad
\beta = \pm\sqrt{\Theta(\theta_0)},
\label{Shadow} 
\end{equation}
where $\Theta(\theta)$ is determined by Eqn~(\ref{Vtheta}). The parameters $\alpha$ and $\beta$ are called the respective horizontal and vertical impact parameters. 

For a remote observer in the equatorial plane of a black hole (with $\theta_0=\pi/2$), the constants $\lambda=L/E$ and $q=\sqrt{Q}/E$ coincide with the respective horizontal and vertical impact parameters in the sky. 

\begin{figure}
	\centering
	\includegraphics[angle=0,width=0.45\textwidth]{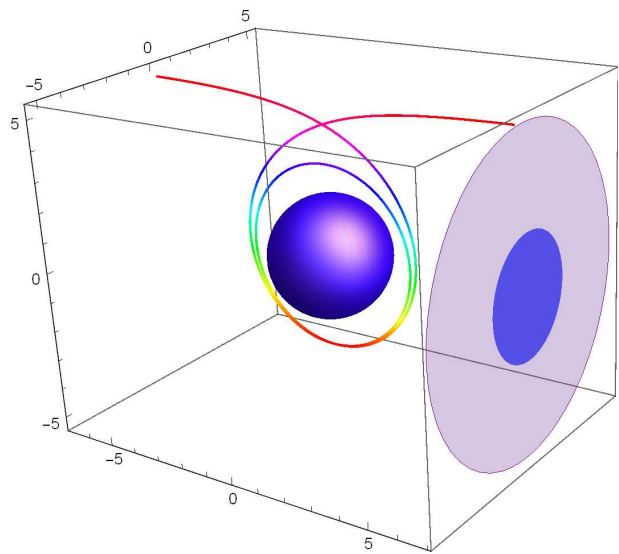}
	\caption{(Color online.) A classical shadow (purple disk) of a Schwarzschild black hole on a remote bright background. Shown is the typical background photon ray (multi-color $3D$ curve) observed near the north pole of the shadow edge. The ray winds up many times around the black hole event horizon (blue sphere) near the radial turning point. Inside the shadow, the blue disk shows a fictive image of the event horizon in the imaginary Euclidean space. }
	\label{fig6}
\end{figure}

Figure~\ref{fig6}  shows the shadow (purple disk) of a Schwarzschild black hole ($a = 0$) with the radius $r_{\rm sh}=3\sqrt{3}\simeq5.2$, as well as the typical $3D$ trajectory of a photon (with the impact parameters $\lambda=0$ and $q=3\sqrt{3}+10^{-3}$) emitted by a remote luminous background. The photon trajectory winds up many times around the black hole horizon (blue sphere) near the radial turning point at $r_{\rm min}\simeq r_{\rm ph}=3$ and reaches the remote observer at the north pole of the black hole shadow. A fictive image of the horizon (blue disk) with the radius $r_{\rm h}=2$ is shown inside the black hole shadow in an imaginary Euclidean space.

\begin{figure}
		\centering 
	\includegraphics[width=0.48\textwidth]{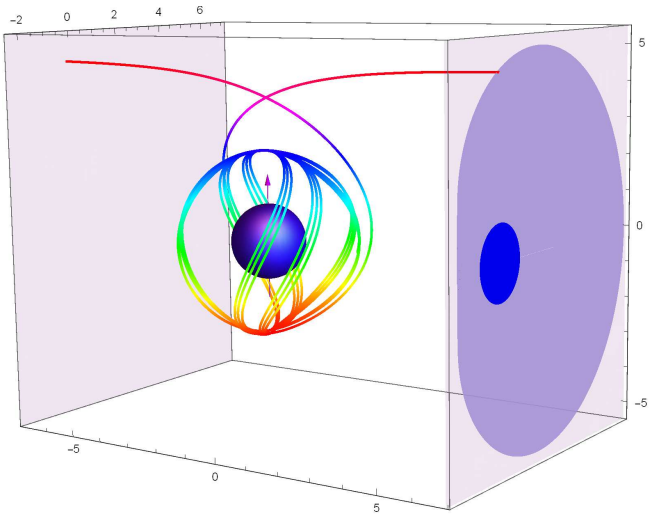}
	\caption{(Color online.) A classical shadow (closed purple region) of an extreme Kerr black hole on a remote bright background. Shown is the typical photon ray (multi-color $3D$ curve) observed near the black hole axis projection crossing with the shadow edge. The ray winds up many times around the black hole event horizon (blue sphere) near the photon sphere. Inside the shadow, the blue disk shows a fictive image of the event horizon in the imaginary Euclidean space. }
	\label{fig7}      
\end{figure}

Figure~\ref{fig7} shows the corresponding shadow from an extremal Kerr black hole $(a=1)$. As a typical example, we present the trajectory of a photon (with orbital parameters $\lambda=0$ and $q=\sqrt{11+8\sqrt{2}}+10^{-3}\simeq4.72$) near the shadow boundary; the photon is emitted by a remote luminous background, winds up many times around the black hole horizon near the radial turning point, and reaches the observer at the crossing point of the shadow edge with the black hole spin axis. The radial turning point of this photon on the light sphere is at $r_{\rm min}=1+\sqrt{2}$. 

 For comparison, in Fig.~\ref{fig5}, we show a numerically calculated direct image, as well as the first and second light echos from a compact star in a circular orbit around the rotating black hole SgrA* as seen by a remote observer in discrete time intervals \cite{doknaz17}. All light echos (multiple images) lie outside the black hole shadow. (See the corresponding numerical animation in \cite{doknaz18b}).

The shadow from an extreme Kerr black hole ($a = 1$) at $\theta_0=0^\circ$ represents a circle with the radius $r_{\rm sh}= 2\sqrt{3+2\sqrt{2}} \simeq4.83$. Photons with orbital parameters $\lambda=0$ and $q=\sqrt{11+8\sqrt{2}}$, which have the radial turning point on the light sphere at $r_{\rm  min}= 1+\sqrt{2}$, produce the contour of this shadow. By numerically solving integral equation~(\ref{eq24a}), it is possible to show that in this case, the event horizon silhouette has a circular form with the radius $r_{\rm eh}\simeq2.035$. 

Shapes of black hole shadows in GR have been studied in many papers. However, presently, as well as in the foreseeable future, observations of real black hole shadows are highly improbable because the radiation intensity of the remote background formed by extended hot gas clouds that are necessary to identify the black hole shadows is far too low to be detected by current interferometric methods. In the case of very bright accreting black holes, their shadows must be blurred by the radiation of hot matter. Here, instead of the black hole shadow, the silhouette of the black hole horizon itself is most likely to be observed. 

The calculation of the gravitational lensing of radiating material during spherical accretion onto a black hole requires that three-dimensional motion of matter falling into the black hole be taken into account, which has not been done yet. However, in the case of a thin accretion disk, the motion of matter is two-dimensional, and the modeling of its gravitational lensing in the gravitational field of a black hole turns out to be quite an easy problem. In Sections 6.2, 6.3, and 7, we describe the features of such gravitational lensing using a simple model of radiation from the inner parts of a thin accretion disk approaching the black hole horizon.

\subsection{Shadow of an accreting black hole}
\label{infall}

Besides a remote luminous background, matter falling into a black hole and radiating near its event horizon can probably be observed. This matter can consist of hot gas, compact gas clouds, or bright compact stars. In the last case, the black hole image would have a shape quite different from the black hole shadow \cite{Luminet79,Bromley97,Fanton97,Fukue03,Fukue03b,Dexter09b,Ru-SenLu16,Luminet19,Shiokawa19,Gucht19}. In particular, in the case of a thin accretion disk, the visible black hole image represents a dark lensed silhouette of its event horizon. The outer boundary of the dark silhouette then represents a lensed equator of the black hole event horizon globe \cite{doknaz19,doknazsm19}.

The model of thin accretion disks around a supermassive black hole describes well the main observational properties of quasars and various types of active galactic nuclei, which are the most powerful sources in the observed universe, and numerous close binary systems with accreting stellar-mass black holes.

We consider the main properties of the event horizon silhouette in GR using model images of supermassive black holes SgrA* and M87* as examples. For this, the most appropriate model is a thin nonselfgravitating accretion disk around the black hole \cite{shaksyn,novthorne73,pageihorne74,Thorne74,Abramowicz13,Yuan14,Lasota15,Zhuravlev15,Compere17}.

\subsubsection{Inner parts of the accretion disk} 
The inner relativistic part of an accretion disk near the event horizon, where most of the energy is released, is the brightest emitting region \cite{Krolik02,Miller07}. A relativistic disk is called geometrically thin if its thickness $h$ is much smaller than the event horizon diameter, $h\ll 2\,r_{\rm h}$.

Circular orbits of particles rotating at radius $r$ in a geometrically thin accretion disk in the equatorial plane of a rotating black hole correspond to the orbital parameters $E$ and $L$ that are determined by the joint solution of equations $R=0$ and $dR/dr=0$, where the effective potential  $R$  is given by Eqn~(\ref{Rr}):
\begin{eqnarray} 
\frac{E}{\mu}&=&\frac{r^{3/2}-2r^{1/2}+a}{r^{3/4}(r^{3/2}-3r^{1/2}+2a)^{1/2}}, \label{Ecirc} \\
\frac{L}{\mu}&=&\frac{r^2-2ar^{1/2}+a^2}{r^{3/4}(r^{3/2}-3r^{1/2}+2a)^{1/2}}.
\label{Lcirc}
\end{eqnarray}
The limit case of a circular orbit is the photon orbit corresponding to infinite $E/\mu$ in Eqn~(\ref{Ecirc}). The radius of the circular  is 
\begin{equation}\label{ISCO}
r_{\rm ph}=2\left\{1+\cos[\frac{2}{3}\arccos(-a)]\right\}.
\end{equation}
In an accretion disk, the inner boundary of stable motion of massive particles is determined by the innermost stable circular orbit (ISCO), $r=r_{\rm ISCO}$ (see \cite{BPT} for more details): 
\begin{equation}\label{ISCO}
r_{\rm ISCO}=3+Z_2-\sqrt{(3-Z_1)(3+Z_1+2Z_2)},
\end{equation}
where
\begin{subequations}
\begin{align}
\label{Z1}  
        Z_1 &=1+(1-a^2)^{1/3}[(1+a)^{1/3}+(1-a)^{1/3}],    \\ 
        Z_2 &=\sqrt{3a^2\!+\!Z_1^2}.
\end{align}
\end{subequations}

The radius of the photon orbit $r_{\rm ph}$ is smaller than $r_{\rm ISCO}$  for any black hole spin $a$. 

By definition, we say that the inner part of a thin accretion disk is the region $r_{\rm h}\leq r\leq r_{\rm ISCO}$ that adjoins the event horizon and contains no stable orbits of the accretion disk matter. This part of the disk is the region of nonstationary accretion. The motion of fragments of matter in this region is fully nonstationary, is controlled by the black hole gravitational field, and is independent of the local properties of matter, including viscosity. In this region, individual fragments of the accretion disk spiral-in independently towards the black hole (see \cite{novthorne73,pageihorne74,Thorne74,Abramowicz13,Yuan14,Lasota15,Zhuravlev15,Compere17,Krolik02,Miller07} for more details).

A widespread misconception is the assumption about rapid decay of emission from the accretion disk in approaching the innermost stable orbit, which is also referred to as the `inner boundary' of the disk. Here, it is erroneously believed that the radiation from the inner disk edge is fully absent or can be ignored. 


We use a simple physically motivated model to describe the nonstationary motion of the emitting material inside the inner region of a thin accretion disk $r_{\rm h}\leq r\leq r_{\rm ISCO}$. In this model, individual disk fragments (as well as compact stars and dense hot gas clouds) move along geodesics with orbital parameters $E$ and $L$  from Eqns~(\ref{Ecirc}) and (\ref{Lcirc}) corresponding to $r=r_{\rm ISCO}$. In other words, separate small disk fragments falling into the black hole `remember' the total energy $E$ and azimuthal angular momentum $L$ they had at the radius $r_{\rm ISCO}$.

To account for gravitational effects only, we also assume that the accretion disk is transparent to radiation. In other words, we entirely disregard scattering and absorption of photons in the plasma around the black hole. The transparency of the hot plasma to radiation down to the event horizon is a necessary condition to observe the black hole image. Correspondingly, the accretion rate onto the black hole should be sufficiently low. This condition is apparently satisfied for SgrA*, which shows shallow activity.
However, for SgrA*, an additional difficulty for observations is a high gas and dust density along the line of sight in the galactic disk. In the case of M87*, which is a high-luminosity object, the scattering and absorption of photons in the surrounding plasma, in principle, can be substantial. In this case, the nonstationary accretion can provide temporary transparency windows. Quite possibly, the team that produced the first observations of M87* was just very lucky to see the black hole when it was transparent to radiation down to the event horizon.

Additionally, in our numerical calculations of the observed radiation from the nonstationary inner parts of the accretion disk, we assume that the energy flux in the comoving frame of small gas fragments is isotropic and is conserved until they reach the gravitational radius $r_{\rm h}$. These model assumptions enable us to calculate the lensed accretion disk brightness quite simply. Importantly, the specific model of radiation from the inner parts of the accretion disc does not affect the shape of the event horizon silhouette and is determined by the gravitational field of a Kerr black hole only.

\begin{figure}
	\includegraphics[width=0.48\textwidth]{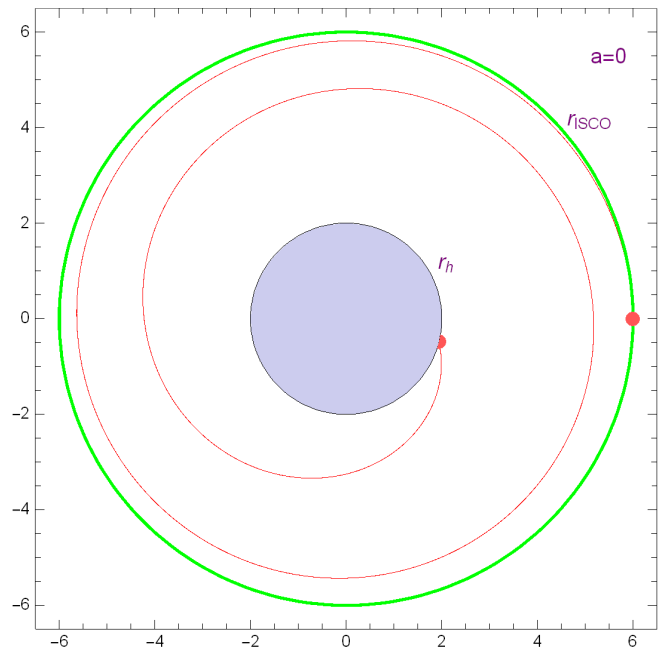}
	\caption{(Color online.) $2D$ trajectory of an accretion disk fragment (small red dot) falling into a Schwarzschild black hole ($a = 0$) in the inner accretion disk part. The fragment spirals in from $r=r_{\rm ISCO}=6$ (green ring) and reaches the event horizon (blue circle) at $r=r_{\rm h}=2$ after making two complete turns.}
	\label{fig8}      
\end{figure}
\begin{figure}
	\includegraphics[width=0.48\textwidth]{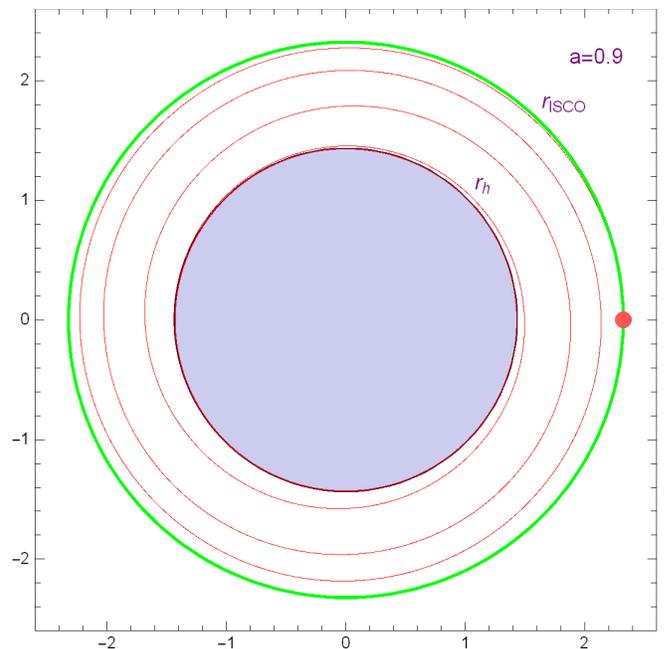}
	\caption{(Color online.) $2D$ trajectory of an accretion disk fragment (small red dot) falling into a rotating black hole with spin $a=0.9$ in the inner accretion disk part. The fragment starts moving from $r=r_{\rm ISCO}\simeq2.32$. It winds up many times on the rotating black hole in approaching the event horizon at $r=r_{\rm h}\simeq1.44$. }
	\label{fig9}      
\end{figure}

Figure~\ref{fig8} shows the two-dimensional trajectory of a small fragment of an accretion disk or a compact gas cloud (clump) falling into a Schwarzschild black hole ($a=0$) in the inner parts of the disk around the black hole event horizon, $r_{\rm h}\leq r\leq r_{\rm ISCO}$. The clump starts moving at $r=r_{\rm ISCO}=6$ (green ring) with the orbital parameters $E/\mu=E(r_{\rm ISCO})/\mu=2\sqrt{2}/3$ and $L/\mu=L(r_{\rm ISCO})/\mu-0.001=\sqrt{3}/2-0.001$, where $E$, $L$ and $r_{\rm ISCO}$ are determined by formulas~(\ref{Ecirc}), (\ref{Lcirc}) and (\ref{ISCO}).

 A similar two-dimensional trajectory for a black hole with spin $a=0.9$ is shown in Fig.~\ref{fig9}. Here, the disk fragment starts falling with the orbital parameters $E/\mu=E(r_{\rm ISCO})/\mu\simeq0.844$ and $L/\mu=L(r_{\rm ISCO})/\mu-0.001\simeq2.09$ at $r=r_{\rm ISCO}\simeq2.32$ (green ring). The disk fragment, in contrast to the case of a Schwarzschild black hole (see Fig.~\ref{fig8}), winds up many times on the rotating black hole in approaching the event horizon at $r=r_{\rm h}\simeq1.44$.

\subsubsection{Gravitational redshift and Doppler effect}

To calculate the energy shift of a photon emitted by matter falling into a black hole inside the inner parts of the accretion disk at $r_{\rm h}\leq r\leq r_{\rm ISCO}$ and detected by a remote observer, we need to take the gravitational redshift and Doppler effects into account. For these calculations, it is convenient to use the orthonormal locally nonrotating frame described in Section~\ref{LNRF}. 

In the LNRF, the azimuthal component of the velocity of a small accretion disk fragment (or a compact gas cloud) at a radius $r$ with orbital parameters $E$, $L$ and $Q=0$ is \cite{BPT,Bardeen70,Bardeen70b} 
\begin{equation}\label{eq2425e}
V^{(\varphi)}=\frac{r\sqrt{\Delta}\,L}{[r^3+a^2(r+2)]E-2aL}.
\end{equation}
The corresponding radial velocity component of the fragment in the LNRF takes the form 
\begin{equation}\label{Vr}
V^{(r)}\!=-\,\sqrt{\frac{r^3\!+\!a^2(r\!+\!2)}{r}}\,
\frac{\sqrt{R(r)}}{[r^3\!+\!a^2(r\!+\!2)]E\!-\!2aL},
\end{equation}
Here, the effective radial potential $R(r)$ is defined by Eqn~(\ref{Rr}) with the parameter $Q=0$. 

We also need expressions for the components of the photon 4-momentum in the LNRF: 
\begin{eqnarray}\label{pphi}
p^{(\varphi)}&=&\lambda\,\sqrt{\frac{r}{r^3+a^2(r+2)}}, \\
p^{(t)}&=&(1-\omega\lambda)\,\sqrt{\frac{r^3+a^2(r+2)}{r\Delta}}, \label{pt} \\
p^{(r)}&=&\frac{1}{r}\sqrt{\frac{(r^2+a^2-a\lambda)^2}{\Delta}-[(a-\lambda)^2+q^2]}.
\label{pr}
\end{eqnarray}
The condition $p^{(i)}p_{(i)}=0$ determines the fourth component of the photon 4-momentum. The energy of a photon in the LNRF is $E_{\rm LNRF}=p^{(t)}$. At the same time, the energy of the same photon in the orthonormal reference frame moving with a velocity $V^{(\varphi)}$ relative to the LNRF is 
\begin{equation}\label{EVphi}
E_{V^{(\varphi)}}=\frac{p^{(t)}-V^{(\varphi)}p^{(\varphi)}}{\sqrt{1-[V^{(\varphi)}]^2}}.
\end{equation}

The photon  energy $E_{V^{(\varphi)}}$ depends only on $\lambda$.  In this frame, the accretion disk fragment still moves with radial velocity
\begin{equation}\label{v}
v=\frac{V^{(r)}}{\sqrt{1-[V^{(\varphi)}]^2}}.
\end{equation}
As a result, the photon energy in the comoving frame of the fragment is 
\begin{equation}
{\cal{E}}(\lambda,q)\!=\!\frac{E_{V^{(\varphi)}}\!-\!vp^{(r)}}{\sqrt{1-v^2}}
\!=\!\frac{p^{(t)}\!-\!V^{(\varphi)}p^{(\varphi)}
	\!-\!V^{(r)}p^{(r)}}{\sqrt{1-[V^{(r)}]^2-[V^{(\varphi)}]^2}}.
\label{calE}
\end{equation}
Correspondingly, the photon energy shift (ratio of the photon frequency detected by a remote observer to the frequency of the same photon in the comoving frame of the fragment) is $g(\lambda,q)=1/{\cal{E}}(\lambda,q)$. In approaching the black hole event horizon, the gravitational redshift starts dominating the Doppler  effect and limit $\lim\limits_{r\to r_{\rm h}} g(\lambda,q)=0$.

We use the photon shift $g(\lambda,q)$ in the Cunningham--Bardeen formalism  \cite{CunnBardeen72,CunnBardeen73} to numerically calculate the energy flux from accretion disk fragments detected by an observer. The results are presented in Figs~\ref{fig14}, \ref{fig15}, \ref{fig17}--\ref{fig20}. The local colors of lensed images of the accretion disc are related to the local black-body temperature in the disk, which is proportional to the energy shift of photons $g(\lambda,q)=1/{\cal{E}}(\lambda,q)$.

In the thin accretion disk model, the brightest point at all spins of the black hole is located at the radius $r=r_{\rm ISCO}$ (see Section~\ref{infall}) (marked with the red star ${\color{red}\star}$ in the corresponding figures). The position of the brightest point on the circle with the radius $r=r_{\rm ISCO}$ corresponds to the photon trajectory without turning points with the maximum admissible azimuthal angular momentum $\lambda>0$ given by the solution of Eqn~(\ref{eq24a}). For a remote observer, this point corresponds to the accretion disk part moving toward the observer with the maximum Doppler factor. 

\subsubsection{Silhouettes of black holes SgrA* and M87*}

In the simplest case of a spherically symmetric Schwarzschild black hole ($a=0$), the boundary of the event horizon image seen by a remote observer is given by the solution of the integral equation 
\begin{equation}
\int_{2}^{\infty}\frac{dr}{\sqrt{R(r)}}
=2\int_{\theta_{\rm min}}^{\pi/2}\frac{d\theta}{\sqrt{\Theta(\theta)}},
\label{a0max}
\end{equation}
Here, $\theta_{\rm min}$ is the turning point in the polar $\theta$-direction on the photon trajectory for the direct image of the probe source found from the equation $\Theta(\theta)=0$. Photons forming the direct image of the source, according to the Cunningham--Bardeed classification scheme of multiple lensed images \cite{CunnBardeen72,CunnBardeen73}, do not cross the black hole equatorial plane along the whole path from the emitter to the observer. The event horizon radius of a Schwarzschild  black hole is $r_{\rm h}=2$, and the turning point is $\theta_{\rm min}= \arccos(q/\sqrt{q^2+\lambda^2})$.  The integral in the right-hand side of (\ref{a0max}) is then equal to $\pi/\sqrt{q^2+\lambda^2}$. As a result, the numerical solution of integral equation (\ref{a0max})  yields the radius of the  image  (silhouette) of the event horizon $r_{\rm eh}=\sqrt{q^2+\lambda^2}\simeq4.457$.  This radius is significantly  smaller than the radius of the black hole shadow $r_{\rm sh}=3\sqrt{3}\simeq5.2$.

In the case of a Kerr black hole ($a\neq0$), the polar turning point (if it exists) is found at
\begin{equation}
\cos^2\theta_{\rm min}\!=\!\frac{\!\sqrt{4a^2q^2\!+\!(q^2\!
+\!\lambda^2\!-\!a^2)^2}\!-\!(q^2\!+\!\lambda^2\!-\!a^2)}{2a^2} 
\label{thetamin} 
\end{equation} 
This expression for $\theta_{\rm min}$ is used for the numerical solution of integral equations (\ref{eq24b}) and (\ref{eq24c}).

\begin{figure}
	\includegraphics[angle=0,width=0.47\textwidth]{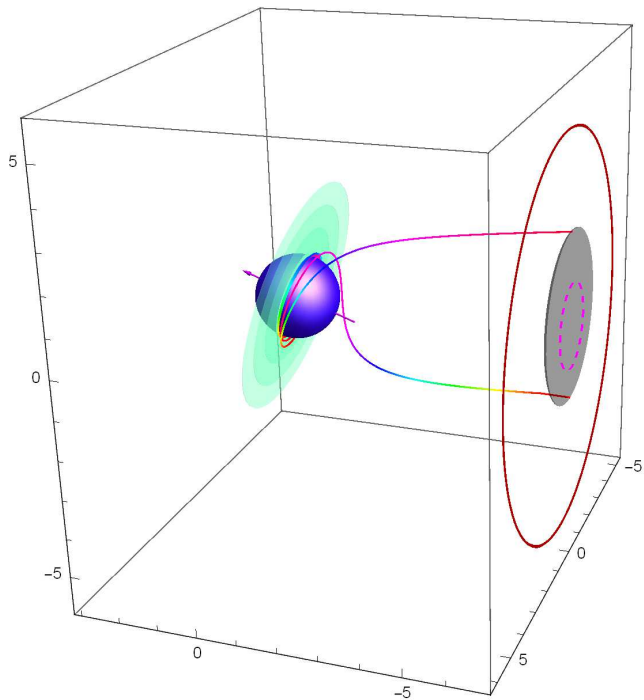} 
	\caption{(Color online.) Two $3D$ trajectories of photons emerging near the event horizon of black hole M87* at different points on a circle of radius $r=1.01\,r_{\rm h}$ in a thin accretion disk (light-green oval) in the equatorial plane of a black hole with spin $a=0.9982$. The rays reach the remote observer near the outer edge of the event horizon silhouette. The closed dark-red curve corresponds to the outer boundary of the black hole shadow.}
	\label{fig10}
\end{figure}

\begin{figure}
	\includegraphics[angle=0,width=0.23\textwidth]{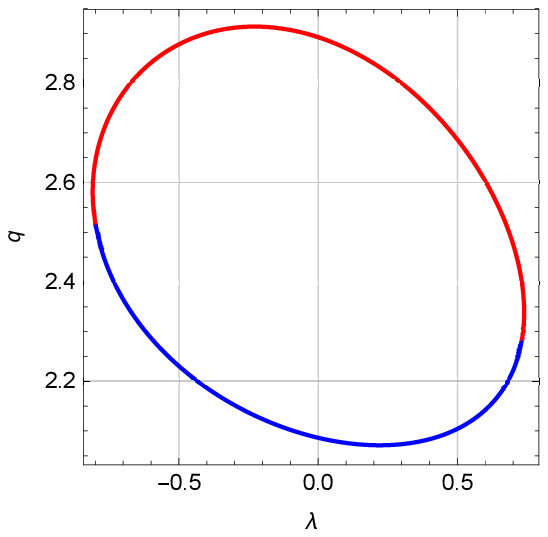}
	\hfill
	\includegraphics[angle=0,width=0.23\textwidth]{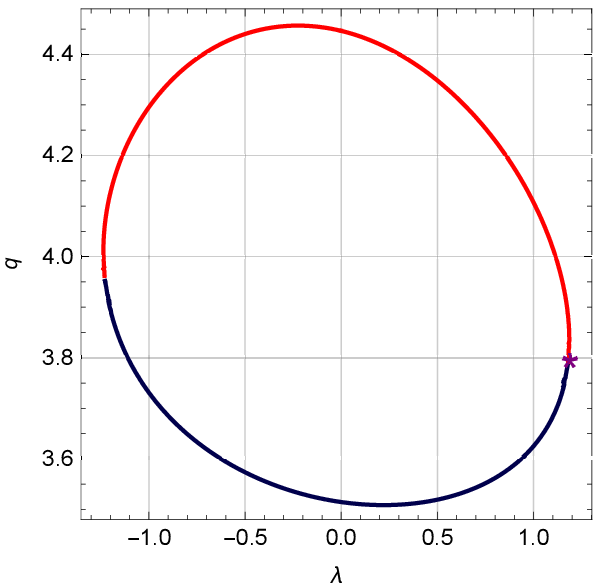}
	\caption{(Color online.) Parameters of photon trajectories  $\lambda$ and $q$ emerging from an accretion disc around the black hole M87* with spin $a=0.75$ at radii (a) $r=1.01\,r_{\rm h}$ and (b) $r=r_{\rm ISCO}$ and reaching the remote observer. The blue color corresponds to photon trajectories without turning points, the red color to rays with a turning point in the polar direction at the polar angle $\theta=\theta_{\rm min}(\lambda,q)$. The red star ${\color{red}\star}$ marks the brightest spot on the accretion disk.}
	\label{fig11}
\end{figure}

The location of the black hole M87* and its accretion disk relative to a remote observer on Earth (or in a near-Earth orbit) is shown in Fig.~\ref{fig10}. The dashed circle in Fig.~\ref{fig10} (and in all other similar figures) marks the black hole horizon image in an imaginary Euclidean space. Figure ~\ref{fig10} also shows $3D$  photon trajectories beginning at different points of the circle with $r=1.01\,r_{\rm h}$ near the event horizon in a thin accretion disk (light-green oval) in the equatorial plane of a black hole with spin $a=0.9982$ and reaching the observer near the outer edge of the event horizon silhouette (grey region).
The orbital parameters of these two photons are $\lambda_1=-0.047$, $q_1=2.19$ and  $\lambda_2=-0.029$, $q_2=1.52$.

Figure~\ref{fig11} shows the plots of parameters $\lambda$ and $q$ of rays emitted from the accretion disk around black hole M87* with spin $a = 0.75$ at radii $r=1.01\,r_{\rm h}$ and $r=r_{\rm ISCO}$ and reaching a remote observer. The blue color in the plots corresponds to photon trajectories without turning points, and the red color to rays with a polar turning point with $\theta=\theta_{\rm min}(\lambda,q)$, which is obtained from the numerical solution of integral equation (\ref{eq24b}). The brightest spot in the accretion disk, marked with the red star ${\color{red}\star}$, is located at the  radius $r_{\rm ISCO} \simeq 1.16\,r_{\rm h}$ and corresponds to the photon orbit parameters $\lambda=1.18$ and $q=3.79$ (or $\alpha=-4.03$ and $\beta=-0.18$).

\begin{figure}
	\includegraphics[width=0.125\textwidth]{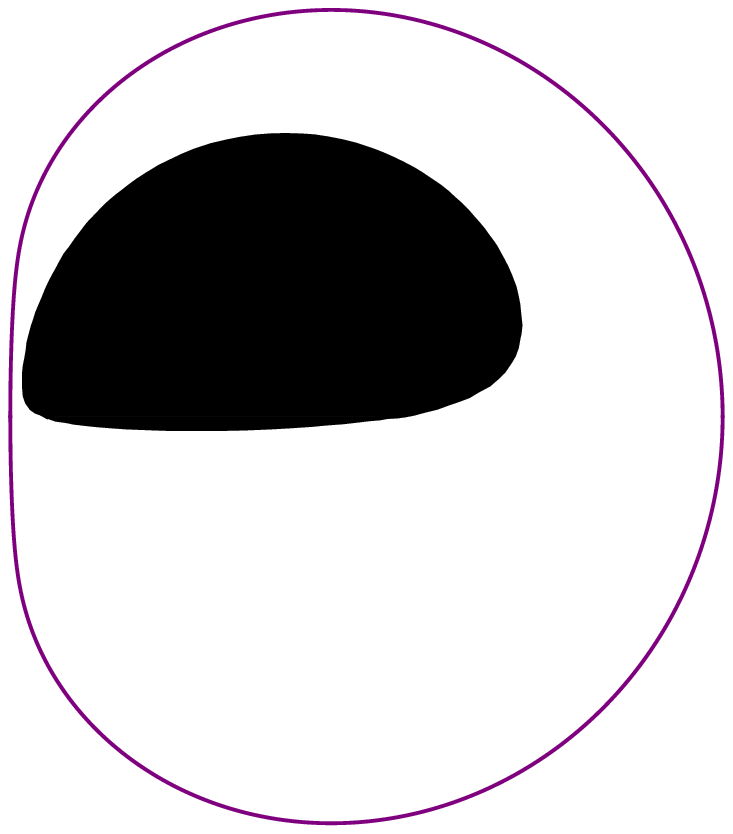}
	\includegraphics[width=0.14\textwidth]{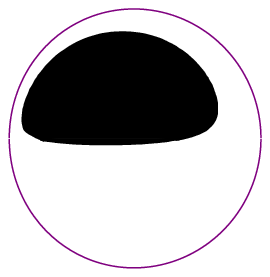}
	\includegraphics[width=0.14\textwidth]{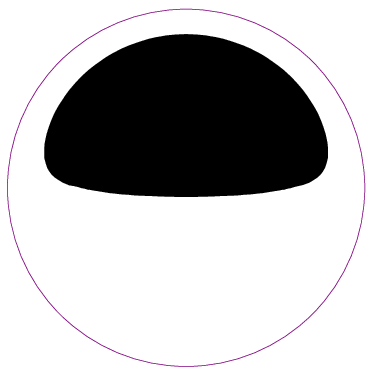}
	\caption{(Color online.) Silhouette of the northern hemisphere of the event horizon of the supermassive black hole SgrA* (black shaded region) inside the black hole shadow (closed purple curve) for the black hole spin (a) $a=0.9982$, (b) $0.65$, and (c) $0$.}
	\label{fig12}      
\end{figure}

The possible shapes of the dark silhouette of the event horizon of the supermassive black hole SgrA* are shown in Fig.~\ref{fig12} for three values of the black hole spin $a$. We note that in the case of the supermassive black hole M87* (taking its spin orientation into account), we observe the silhouette of its southern hemisphere (black shaded region) located inside the corresponding shadow of this black hole (closed purple curve). 

The supermassive black hole SgrA* in the galactic center has the mass $M= (4.3 \pm 0.3)\times 10^6M_\odot$, which is three orders of magnitude smaller than that of the supermassive black hole M87*, but SgrA* is three orders of magnitude closer than M87*. Therefore, the event horizons of these black holes have about the same angular sizes accessible to EHT observations. The spin axis of SgrA* is almost certainly close to the Milky Way rotational axis  \cite{Psaltis15}. 
For certainty, we assume that the remote observer lies close to the black hole equator, more precisely, with $\cos\theta_0=0.1$ or $\theta_0\simeq84.24^\circ\!$. The possible forms of the event horizon silhouette of the supermassive black hole SgrA* are shown in Fig.~\ref{fig12} for three values of the black hole spin $a$.

\begin{figure}
	\includegraphics[width=0.48\textwidth]{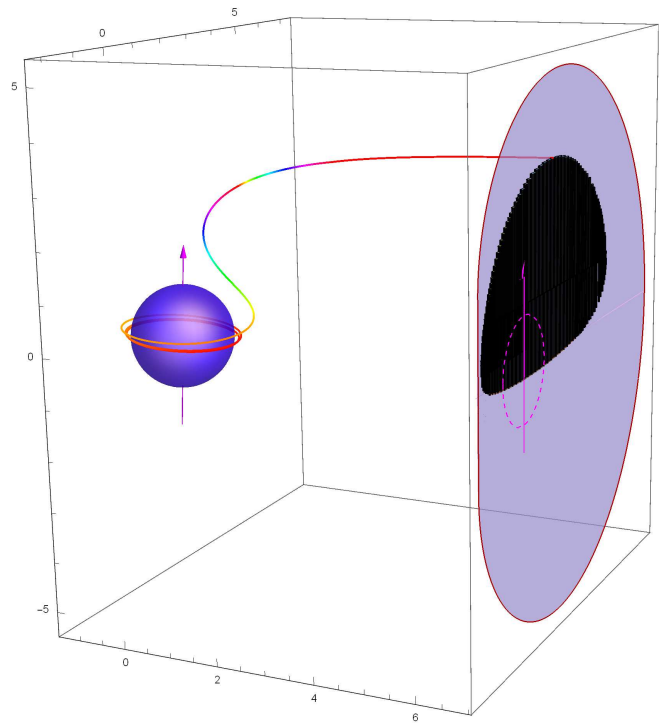}
	\caption{(Color online.) The dark silhouette of the northern hemisphere of the black hole horizon (black shaded region) in the case of a thin accretion disk in the equatorial plane of a black hole with spin $a=0.9982$  corresponding to SgrA* in the galactic center. The contour of this silhouette is formed by strongly high-redshifted photons emitted near the event horizon by hot accretion disk gas and registered by the remote observer. One such photon ray is presented.}
	\label{fig13}      
\end{figure}

\begin{figure}
	\includegraphics[width=0.47\textwidth]{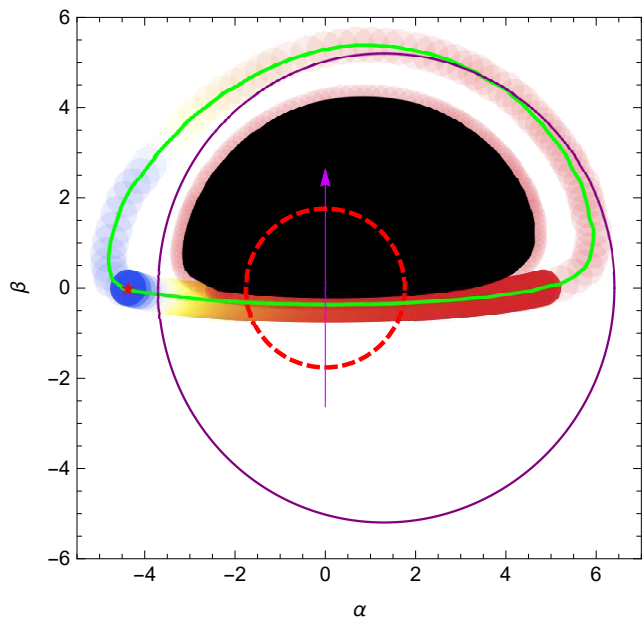}
\caption{(Color online.) Lensed image of the inner accretion disk adjoining the event horizon of the black hole SgrA* with spin $a=0.65$. The closed green curve is the image of the equatorial circle with the radius $r_{\rm ISCO}\simeq 3.25 \simeq 1.85r_{\rm h}$. Shown is the bright emission from a thin accretion disk ring at the radius $r_{\rm ISCO}$ and faint emission from a thin ring at the radius $1.01\,r_{\rm h}$ suppressed by a significant gravitational redshift near the event horizon. The brightest spot on the accretion disk is marked with the red star (also in other similar figures). The black shaded region shows the silhouette of the event horizon northern hemisphere. The outer boundary of this silhouette is the gravitationally lensed image of the event horizon equator. The closed purple line marks the black hole shadow boundary. A sharp brightness change on the lensed image of the thin ring at the radius $r_{\rm ISCO}$ is due to the transition from light rays without turning points (numerical solution of Eqn~(\ref{eq24a})) to those with the turning point $\theta_{\rm min}$ (numerical solution of Eqn~(\ref{eq24b})). The latter rays reach the observer along longer paths and provide much lower local brightness than those without turning points.}
	\label{fig14}      
\end{figure}

In Fig.~\ref{fig13}, the dark silhouette of the northern hemisphere of the event horizon is visible (black shaded region) in the case of a thin accretion disk in the equatorial plane of a black hole with the spin $a=0.9982$  corresponding to SgrA*. 
Strongly redshifted photons emitted near the event horizon by hot  matter and reaching the remote observer form the contour of this silhouette. Figure~\ref{fig13} shows one of these light rays with the orbital parameters $\lambda=0.063$ and $q=0.121$. The photon starts in the black hole equatorial region at the radius $r=1.01\,r_{\rm h}$ and is observed near the outer edge (contour) of the silhouette of the event horizon northern hemisphere. Similar silhouettes for other spins $a$ are shown in Figs~\ref{fig14} and ~\ref{fig15} for SgrA* and M87*.

Black hole silhouettes with shapes very similar to those shown in Figs~\ref{fig12}--\ref{fig15} have been numerically calculated for many years from accretion disk images around black holes (see, e.\,g., \cite{Luminet79,Bromley97,Fanton97,Fukue03,Fukue03b,Dexter09b,Ru-SenLu16,Luminet19,Shiokawa19,Gucht19,White20}).
However, these studies have not identified these silhouettes with a hemisphere of the black hole event horizon. A principal feature of these numerical models has been accounting for radiation from the inner accretion disk at $r_{\rm h}\leq r\leq r_{\rm ISCO}$ when a dark image of the northern or southern hemisphere of the event horizon globe (depending on the viewing angle of the black hole spin axis) is observed inside the accretion disk, with the outer edge of this image being the lensed image of the equator of the event horizon globe. Here, the size of the event horizon silhouette is significantly smaller than the expected black hole shadow diameter. 

Model images of accretion disks around a Schwarzschild black hole ($a=0$) with the inner disk boundary at $r_{\rm ISCO}=6$ much exceeding the circular photon orbit radius $r_{\rm ph}=3$ have also been elaborated. In this case, the accretion disk without radiation at $r<r_{\rm ISCO}$ is, for a black hole, a replica of the remote background. Therefore, these models reproduce the black hole shadow and not the event horizon silhouette (see, e.\,g., ~\cite{Gyulchev19,Shaikh19b,Feng19}).

We also note that in the famous supercomputer simulation for the movie \textit{Interstellar}, radiation from the inner accretion disk ($r_{\rm h}\leq r\leq r_{\rm ISCO}$) approaching the black hole event horizon was ignored at the producer's request. As a result, the gravitational lensing of the stationary part of the model accretion disk with the inner boundary at $r=9.26$ around a black hole with the spin $a=0.999$ yielded a dark image of the black hole shadow but not the dark silhouette of  the event horizon. The producer also demanded that the energy redshift of photons from the accretion disk be taken into account to model the observed disk brightness but not the color \cite{Thorne15,Luminet15,Luminet18}. As a result, this numerical supercomputer simulation also relates to models that effectively reproduce the black hole shadow image only, ignoring possible disk radiation near the event horizon.

\subsubsection{Brightest spot on the accretion disk}

\begin{figure}
	\makebox{\includegraphics[width=3.4cm,trim=0cm -1.9cm 0cm 0cm]{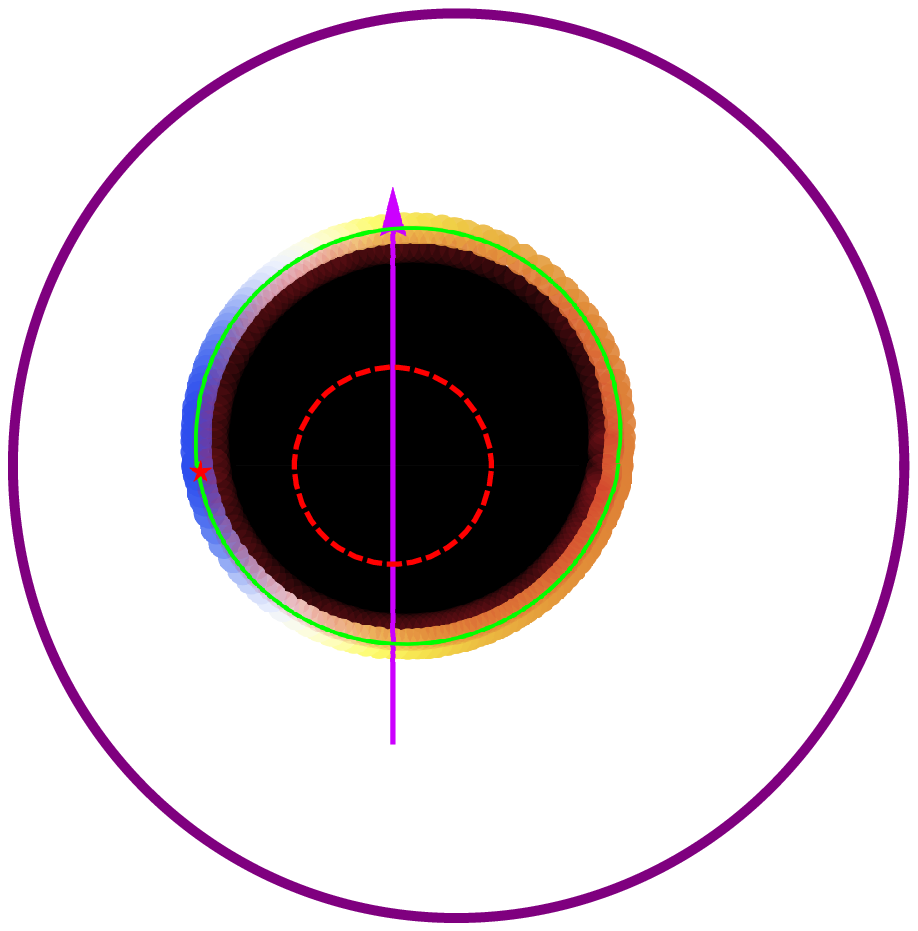}} 	\hfill
	\mbox{\includegraphics[width=5.1cm,keepaspectratio]{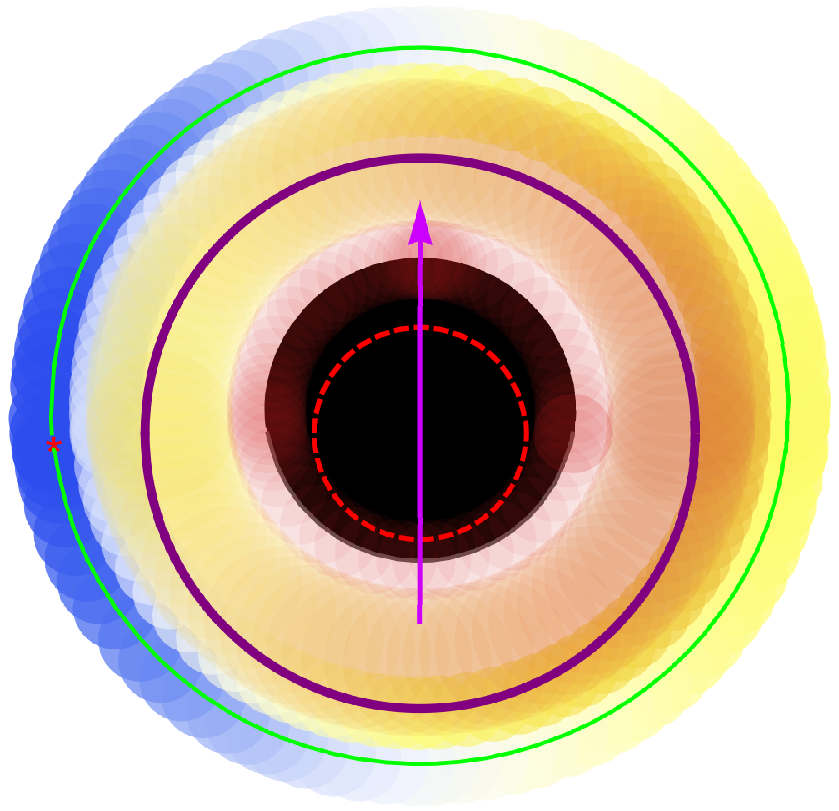}}
	\caption{(Color online.) Radiation from the inner accretion disk part in the equatorial plane of the black hole M87* and the silhouette of the southern hemisphere of the event horizon (black shaded region) located inside the black hole shadow (closed purple curve) for the black hole spins (a) $a=0.9882$ and (b) $a=0$.}
	\label{fig15}
\end{figure}

In relativistic models of thin accretion disks, the local disk brightness measured by a remote observer increases in approaching the black hole \cite{shaksyn,novthorne73,pageihorne74,Thorne74,Abramowicz13,Yuan14,Lasota15,Zhuravlev15,Compere17}. The intensity is thought to reach a maximum at the radius $r_{\rm ISCO}$ (\ref{ISCO}). In approaching a black hole, the redshift of emitted photons registered by the remote observer increases. However, an increase in the local brightness at $r<r_{\rm ISCO}$, in principle, could be sustained by Doppler beaming in the rapidly rotating disk part directed towards the observer. Our numerical simulations \cite{doknaz19b} do not support this hypothesis. In the inner parts of a thin accretion disk (i.\,e., at $r_{\rm h}\leq r\leq r_{\rm ISCO}$) adjoining the black hole horizon, the gravitational redshift of photons dominates over the Doppler effect at any black hole spin $a$.

In our numerical simulations, we use the physically justified assumption that the nonstationary fall of small accretion disk fragments into a black hole at $r_{\rm h}\leq r\leq r_{\rm ISCO}$ occurs along geodesics with orbital parameters $E$ and $L$ from Eqns~(\ref{Ecirc}) and (\ref{Lcirc}) with $r=r_{\rm ISCO}$. In other words, we postulate that small fragments of the accretion disk falling into the black hole `remember' their orbital parameters at the radius $r_{\rm ISCO}$. The numerical modeling enables us to find the location of the brightest spot on the ring with the radius $r_{\rm ISCO}$. This spot corresponds to the direct-image photons (those reaching a remote observer without crossing the equatorial plane of the black hole), which have a maximum possible azimuthal angular momentum $\lambda$. For such a photon, the Doppler boost is maximal. In particular, the brightest spot on the accretion disk around the black hole SgrA* with spin $a=0.65$ at the radius $r_{\rm ISCO}$ is determined by the direct-image photon ray with parameters $\lambda=4.29$ and $q=0.430$ corresponding to the impact parameters $\alpha=-4.32$ and $\beta=-0.042$ in the sky (see Fig.~\ref{fig16} in Section~\ref{cartography}).

In Figs~\ref{fig14}, \ref{fig15}, \ref{fig18} and \ref{fig20}, we show examples of the lensed emission from the nonstationary inner parts of an accretion disk $r_{\rm h}\leq r\leq r_{\rm ISCO}$ adjoining the event horizon $r_{\rm h}$, and producing a direct image of the event horizon with the photon energy gravitational redshift and Doppler effect taken into account as $g(\lambda,q)=1/{\cal{E}}(\lambda,q)$, where ${\cal{E}}(\lambda,q)$ is given by formula (\ref{calE}). The visible image of a black hole surrounded by a moderately thick luminous accretion disk with thickness $h\leq2\,r_{\rm h}$ includes a dark silhouette in the near-polar region of the event horizon globe bounded by a parallel with the latitude angle $\arcsin[h/(2\,r_{\rm h})]$. For $h\to0$ (a thin accretion disk), a dark silhouette of the northern hemisphere of the event horizon of the black hole illuminated by the inner part of the accretion disk must be observed. The contour of this dark silhouette is the lensed image of the equator on the event horizon globe.

\subsection{Mapping of the event horizon}
\label{cartography}

\begin{figure}
	\centering 
	\includegraphics[angle=0,width=0.45\textwidth]{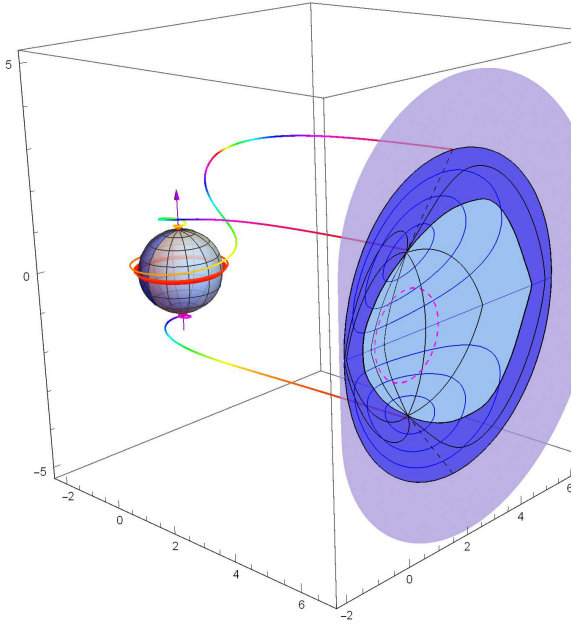}
	\caption{(Color online.) Reconstructed image of the event horizon of an extreme Kerr black hole ($a=1$) and the individual photon trajectories (multi-color $3D$ curves) forming this image and reaching a remote observer in the black hole equatorial plane. The photons come from the vicinity of the north and south poles and equator of the event horizon globe. Shown are some parallels (closed blue curves) and meridians (black curves) on the event horizon globe (blue sphere) and its sky projection (blue shaded region). The dark blue part of the image is the sky projection of the rear hemisphere of the event horizon.}
	\label{fig16}
\end{figure} 

\begin{figure}
	\centering
	\includegraphics[width=.45\textwidth,origin=c,angle=0]{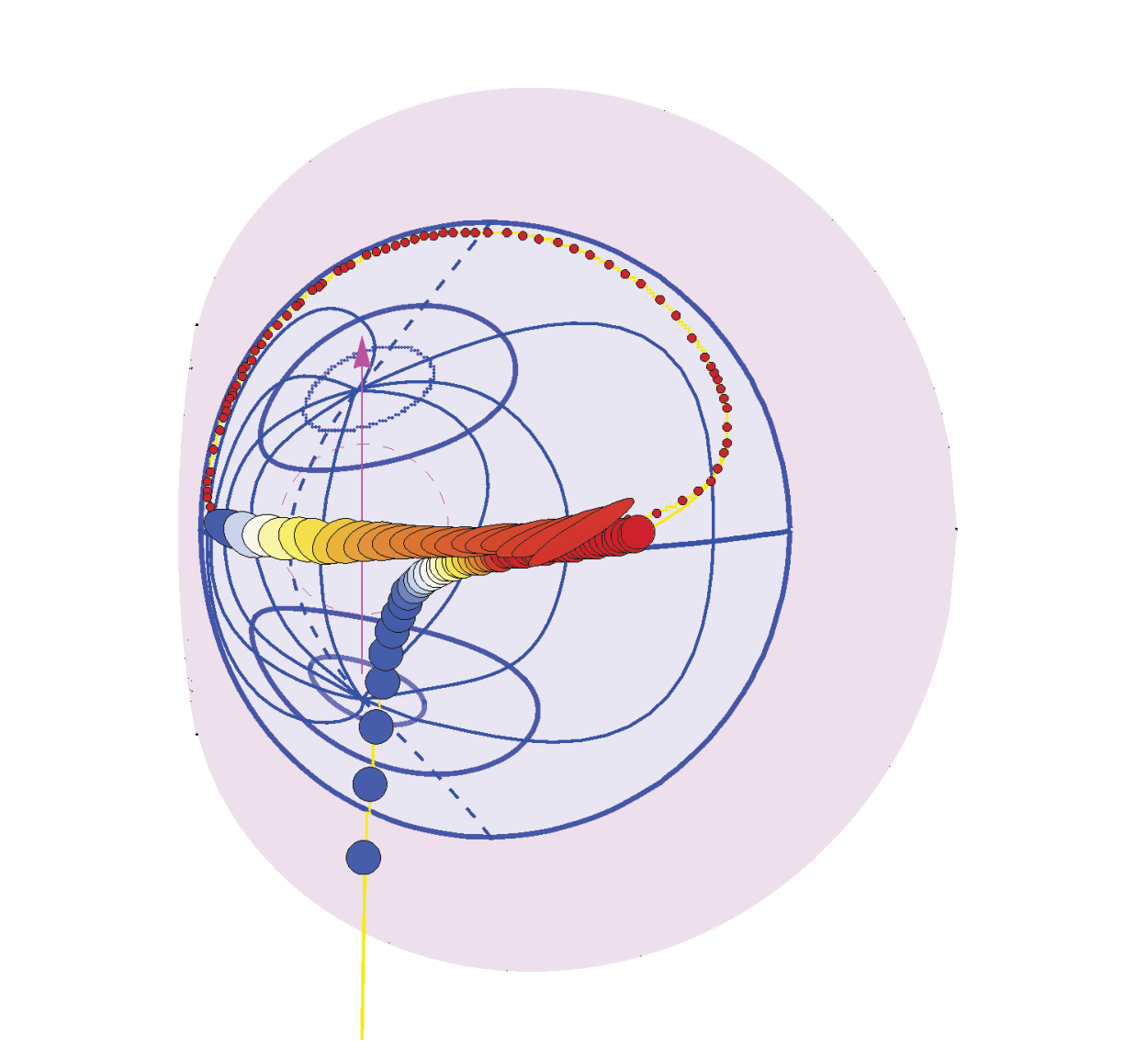}
	\caption{(Color online.) Numerical modeling of the gravitational lensing of a compact star falling into the rotating black hole SgrA* (with the assumed spin $a=0.9982$) observed from infinity in discrete time intervals. The star, with zero azimuthal angular momentum, moves in the black hole equatorial plane. In approaching the event horizon, the lensed images of the source fall inside the black hole shadow (light purple region) and then start winding up the black hole by gradually approaching the equator $\theta=\pi/2$ in the lensed image of the event horizon globe (light blue region). Shown is the first cycle of this winding-up.}
	\label{fig17}
\end{figure}

\begin{figure}
	\includegraphics[width=0.48\textwidth]{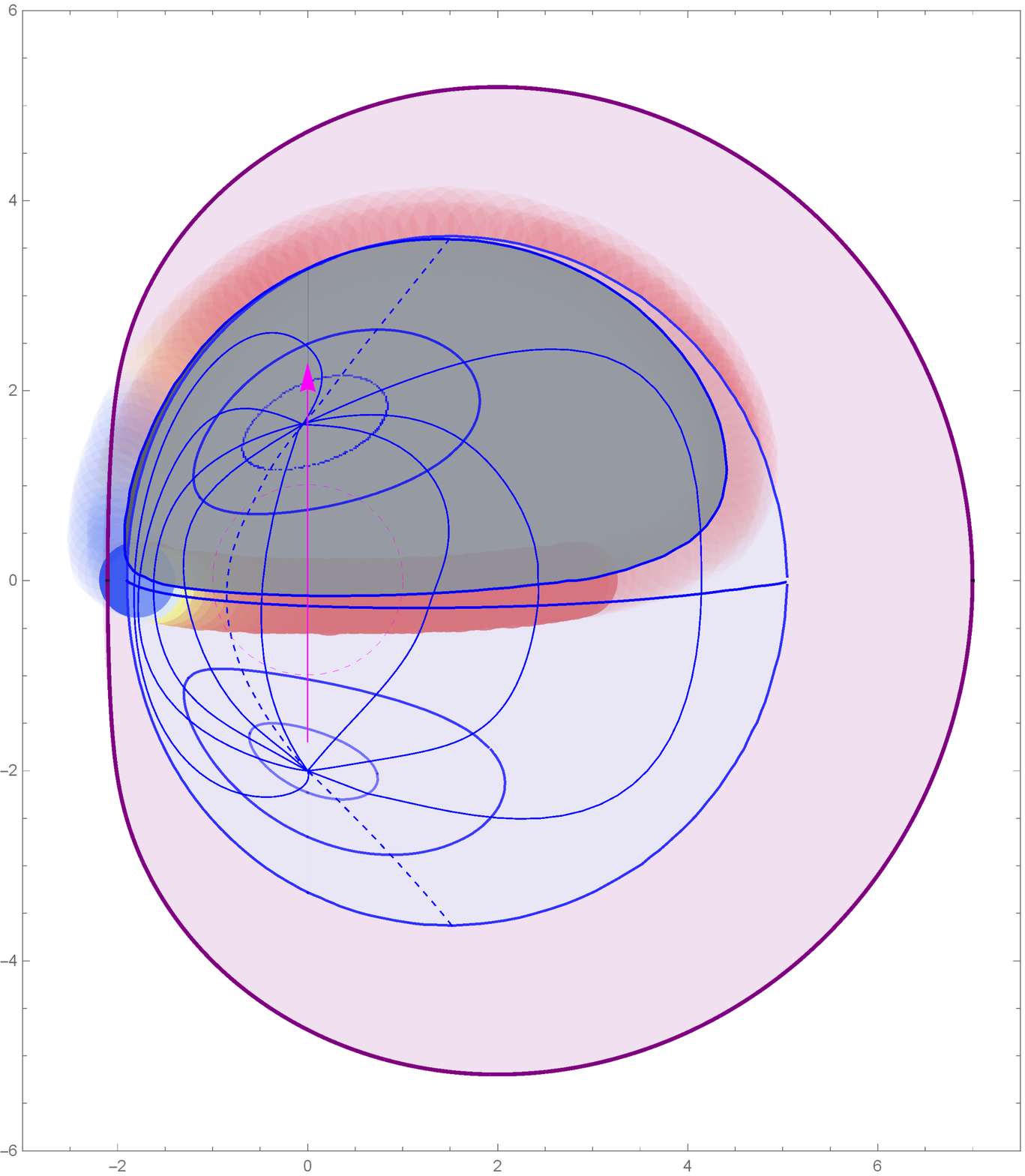}
	\caption{(Color online.) Complete map of the event horizon globe (closed blue region with a dark spot) as recovered from observations of compact radiating sources falling into a black hole from different sides. The blue curves show parallels and meridians on the restored image of the event horizon globe. The vast bounded light purple region is the shadow of the black hole SgrA* with the spin $a=0.9982$. In the image of the black hole surrounded by a moderately thick radiating accretion disk with a thickness  $h\leq2\,r_{\rm h}$, a dark silhouette of the near-polar region of the event horizon globe above the parallel with the latitude angle $\arcsin[h/(2\,r_{\rm h})]$ is visible. For $h\to0$ (a thin accretion disk), a dark silhouette of the northern hemisphere of the event horizon of this black hole (dark grey region) illuminated by the inner accretion disk with color changing from blue to red is to be observed. The contour of this dark silhouette is the lensed image of the event horizon equator.}
	\label{fig18}      
\end{figure}

We now proceed from describing the dark silhouette of the event horizon to mapping (or reconstructing) the image of the entire event horizon globe.

It is convenient for what follows to do a thought experiment by throwing emitting probes from all sides into a black hole (e.\,g., neutron stars or compact emitting hot gas clouds). The lensed image of the probes is registered and studied by a remote observer. When approaching the black hole, the radiation from the probes is observed increasingly redshifted, up to infinity on the horizon. The sky location of every `last' photon from the probe crossing the horizon, which can still be detected by the remote observer, provides information on the unique specific point on the black hole horizon.

Measurements of the last-detected photons from numerous radiating probes falling from all sides into the black hole enable one not only to `see' the outer boundary (contour) of the lensed dark silhouette of the event horizon, but also to reconstruct (or `map') the image of the entire event horizon globe (see \cite{Dokuch19,doknazsm19,doknaz19b} for more details). The numerical results of event horizon mapping are in full agreement with analytic mapping \cite{gralla19}.

Formally speaking, photons emitted from the vicinity of the event horizon and detected by the remote observer enable her/him to uniquely project the whole event horizon globe onto a bounded sky region. We refer to the result of such a mapping as a `lensed image' or, simply, an `image' of the event horizon. On this `image,' the event horizon can be `seen' from all sides simultaneously.

Photons emitted near the event horizon by the luminous matter falling into a black hole and detected by the remote observer are strongly redshifted. Therefore, the accuracy of the `mapping' of the black hole event horizon depends on the detector sensitivity. The lensed image of the entire event horizon is entirely inside the black hole shadow in the sky.

Figure~\ref{fig16} shows a map (reconstruction) of the event horizon of an extreme Kerr black hole ($a=1$) and the typical light rays forming this image as seen by a remote observer in the equatorial plane of the black hole. The photons are emitted from the northern and southern poles of the event horizon globe with orbital parameters $\lambda=0$, $q = 1.77$, as well as from its equator with $\lambda=-1.493$ and $q=3.629$. The largest purple shaded region is the black hole shadow. Shown are some parallels (blue looped curves) and meridians (black curves) on the event horizon globe (blue sphere) and its sky projection (blue shaded region). The light blue part of the image is the sky projection of the front hemisphere of the event horizon globe. Correspondingly, the dark-blue part of the image is the sky projection of the rear region of the event horizon globe \cite{doknaz19,Dokuch19}. 
The numerical modeling results shown in Fig.~\ref{fig17} demonstrate an example of the gravitational lensing of a massive spherical radiation source (a compact star or gas cloud) with the trajectory parameters $\gamma=1$, $\lambda=q=0$ that falls into a black hole with the spin $a=0.9982$ in its equatorial plane. The remote observer lies at the latitude with $\cos\theta=0.1$. In approaching the event horizon, the lensed images of the infalling source are projected on the sky inside the black hole shadow and start `winding up' many times on the black hole, with each winding approaching the equatorial parallel $\theta=\pi/2$ on the event horizon globe. The lensed image of this radiation source identifies (maps) the event horizon equator $\theta=\pi/2$ asymptotically in time (after many turns around the black hole). (See the animation of this process in \cite{doknaz18c}). 

Figure~\ref{fig18} shows the black hole shadow from SgrA* with the spin $a=0.9982$ and the silhouette of the northern hemisphere of its event horizon illuminated by the inner part of a thin accretion disk with a thickness h much smaller than the event horizon diameter, $h\ll2\,r_{\rm h}$. If the radiating accretion disk is moderately thick, $h \leq 2\,r_{\rm h}$, the dark spot on the corresponding image of the black hole is the silhouette of the polar part of the event horizon globe above the parallel with latitude angle $\arcsin[h/(2\,r_{\rm h})]$. Here, the intermediate region of the event horizon globe between this latitude and the equator is fully or partially blurred by the emission from the inner accretion disk. If, conversely, the accretion near the event horizon is very thick, $h \geq 2\,r_{\rm h}$, the dark spot on the black hole image is either absent or partially smeared out by the inner accretion disk emission.

\section{Spin of the supermassive black hole M87*}

\begin{figure}
	\includegraphics[width=0.49\textwidth]{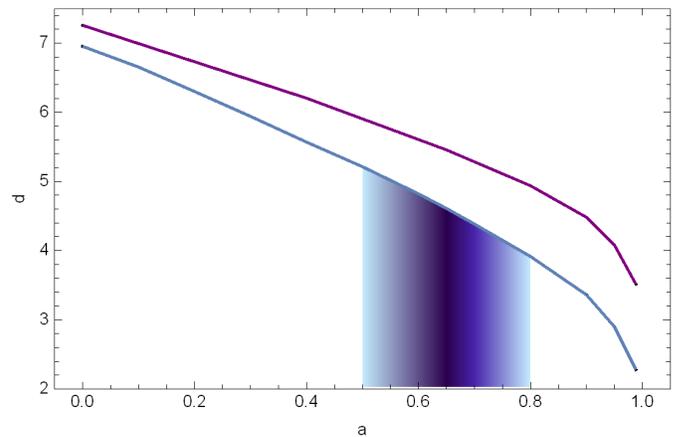}
	\caption{(Color online.) Distance $d$ (in units $GM/c^2$) between the brightest point on a thin accretion disk lying in the black hole equatorial plane and the center of the visible silhouette of the event horizon as a function of the spin $a$ for the black holes M87* and SgrA* (lower and upper curves, respectively). Shaded is the region $\pm1\sigma$ around the spin $a=0.75$. The spin $a=0.75$ corresponds to the best-fit accretion disk model and observed location of the bright spot on the M87* image shown in Fig.~\ref{fig20}.}
	\label{fig19}
\end{figure}

The model of the lensed image of a part of a thin accretion disk, $r_{\rm h}\leq r\leq r_{\rm ISCO}$, considered in Section~\ref{infall} allows us to find the dependence of the distance $d$ of the brightest spot on the accretion disk to the dark silhouette center as a function of the black hole spin $a$. The dependences  $d(a)$ are plotted in Fig.~\ref{fig19}  for supermassive black holes SgrA* and M87*. The application of the function  $d(a)$ to EHT image interpretation suggests that the thin accretion disk model best fits the image of the supermassive black hole M87* when its spin is $a=0.75 \pm 0.15$ (with a $1\sigma$ statistical error) \cite{doknaz19b}. This estimate is in agreement with independent evaluations of the M87* spin \cite{Broderick08,Li09,Feng17,Sobyanin18,Nokhrina19,Tamburini19,Bambi19,Nemmen19,Nalewajko19}.

\begin{figure}
	\includegraphics[width=0.49\textwidth]{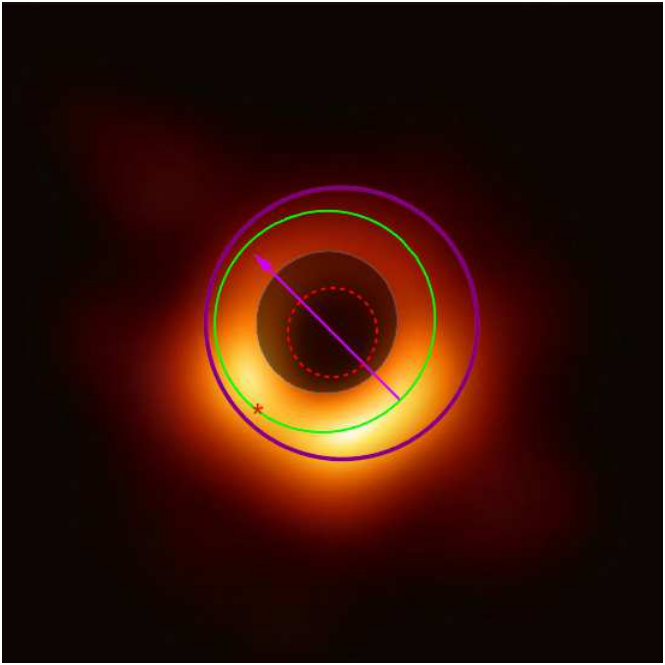}
	\caption{(Color online.) Superposition of the EHT image of the black hole M87* and the thin accretion disk model for the black hole spin $a=0.75$. The purple arrow marks the black hole spin axis. The small dashed circle corresponds to the black hole event horizon in the imaginary Euclidean space. The green closed curve is the lensed image of a ring with the radius $r=r_{\rm ISCO}$. The brightest spot on the accretion disk is marked with the red star ${\color{red}\star}$. The largest closed purple curve is the boundary of the black hole shadow, which is invisible in this image. The black spot at the image center is the observed black hole event horizon silhouette predicted by GR. (See \cite{doknaz19b} for more details).}
	\label{fig20}
\end{figure}

Figure~\ref{fig20} shows the superposition of the EHT M87* image with the thin accretion disk model for the black hole spin $a=0.75$. The brightest point on the accretion disk marked in Fig.~\ref{fig20} by the red star ${\color{red}\star}$  lies on a circle with the radius $r_{\rm ISCO}$ between two bright spots on the M87* image. The presence of two luminous spots instead of one bright spot is due to an insufficient number of Fourier components used to restore the interferometric image observed by eight telescopes. The dark spot at the image center is the black hole horizon silhouette predicted by GR. The black hole shadow, which is much larger than the event horizon, is not visible in this image.

\section{Conclusion}

Possible shapes of dark images (shadows) of black holes depend sensitively on the distribution of the surrounding luminous matter and the black hole spin viewing angle. 

In the case of a remote radiating background behind the event horizon of a black hole, a classical dark shadow of the black hole with maximum size can be observed. A shadow with a minimum diameter can be seen if the black hole is illuminated by the inner parts of an accretion disk adjoining the event horizon. If the accretion disk is thin, the contour (the outer edge) of this shadow represents the lensed image of the event horizon equator. 

The dark silhouette of the southern hemisphere of the event horizon is visible in the M87* image obtained by the EHT collaboration with a record high angular resolution. It is the first direct proof of the existence of black holes because, in the M87* image, on top of the bright background, a dark spot is seen whose size is of the order of the event horizon of this supermassive object. Satisfaction of exactly this condition is direct proof that black holes exist in the Universe. The size and form of the dark spot in the M87* images fully agree with GR predictions for the expected black hole silhouette.

There is a fundamental possibility of reconstructing the lensed image of the entire event horizon globe by observing compact luminous objects falling into  black hole outside its equatorial plane. The reconstructed  `image' of the event horizon is projected on the sky inside he classical black hole shadow.  An analogous statement on the possibility of mapping the entire event horizon globe of a Schwarzschild black hole is made in \cite{Schwar}.

Photons emitted by objects falling into a black hole close to its horizon are very strongly redshifted for a remote observer. Therefore, the accuracy of determination of the sky position of the last detected photon from an object falling into the black hole and hence the accuracy of the event horizon mapping is determined by the ability of a telescope to detect low-energy photons.

We also note that gravitational-wave bursts registered by laser interferometers \cite{LIGO16a,LIGO16b,LIGO16c,LIGO16d,LIGO16e,Scheel14,Cher16,Cher16b,Reitze17,Postnov19} can be explained only by coalescences of stellar-mass compact objects with the proper size: of the order of their event horizons. From the GR standpoint, only black holes or neutron stars can be such objects. Nevertheless, gravitational wave detection provides reliable proof of the existence of black holes, but only, in fact, indirectly, because it does not prove that the proper sizes of the coalescing objects are smaller than their event horizons. 

\begin{acknowledgements}

 The authors thank Eugeny Babichev, Victor Berezin, Yury Eroshenko and Alexei Smirnov for the fruitful discussions and critical remarks. The authors acknowledge financial support from the Russian Foundation for Basic Research grant no. 18-52-15001-NCNIa.
\end{acknowledgements}

\end{document}